\documentclass[sigconf,10pt]{acmart}

\usepackage{tikz}
\usepackage{xspace}
\usepackage{balance}
\usepackage{flushend}
\usepackage{amsmath}
\usepackage{bm}
\usepackage{algorithmic}
\usepackage{graphicx}
\usepackage{caption}
\usepackage{gensymb}
\usepackage{subfigure}
\usepackage{textcomp}
\usepackage{xcolor}
\usepackage{colortbl}
\usepackage{booktabs}
\usepackage{url}
\usepackage{multirow}

\AtBeginDocument{%
  }

\copyrightyear{2025}
\acmYear{2025}
\setcopyright{cc}
\setcctype{by}
\acmConference[ACM MOBICOM '25]{The 31st Annual International Conference on Mobile Computing and Networking}{November 4--8, 2025}{Hong Kong, China}
\acmBooktitle{The 31st Annual International Conference on Mobile Computing and Networking (ACM MOBICOM '25), November 4--8, 2025, Hong Kong, China}
\acmDOI{10.1145/3680207.3723462}
\acmISBN{979-8-4007-1129-9/2025/11}

\newcommand{\SystemName}{QuinID\xspace}
\newcommand{\SystemNameNoSpace}{QuinID}
\newcommand{\SystemMaster}{QuinReader\xspace}
\newcommand{\SystemSlave}{QuinTag\xspace}

\newcommand{\revision}[1]{#1}

\begin{document}
\title{\SystemNameNoSpace: Enabling FDMA-Based Fully Parallel RFID with Frequency-Selective Antenna}

\author{Xin Na$^1$, Jia Zhang$^1$, Jiacheng Zhang$^1$, Xiuzhen Guo$^2$, Yang Zou$^1$, Meng Jin$^3$\\Yimiao Sun$^1$, Yunhao Liu$^1$, Yuan He$^1$}
\authornote{Yuan He is the corresponding author.} 
\affiliation{
  \institution{$^1$Tsinghua University, $^2$Zhejiang University, $^3$Shanghai Jiao Tong University}
  \city{}
  \country{}
}
\email{{nx20, j-zhang19, zhangjc21, zouy23, sym21}@mails.tsinghua.edu.cn,}
\email{guoxz@zju.edu.cn, jinm@sjtu.edu.cn, {yunhao, heyuan}@tsinghua.edu.cn}


%----------------------- ACM Reference Format & Copyright Removal -----------------
% \settopmatter{printacmref=false} % Removes citation information below abstract
% \renewcommand\footnotetextcopyrightpermission[1]{} % removes footnote with conference information in first column

%------------------------------------ Abstract -------------------------------------
% !TeX root = ../main.tex

\begin{abstract}
Parallelizing passive Radio Frequency Identification (RFID) reading is an arguably crucial, yet unsolved challenge in modern IoT applications. Existing approaches remain limited to time-division operations and fail to read multiple tags simultaneously. In this paper, we introduce \SystemName, the first frequency-division multiple access (FDMA) RFID system to achieve fully parallel reading. We innovatively exploit the frequency selectivity of the tag antenna rather than a conventional digital FDMA, bypassing the power and circuitry constraint of RFID tags. Specifically, we delicately design the frequency-selective antenna based on surface acoustic wave (SAW) components to achieve extreme narrow-band response, so that \SystemName tags (i.e., \SystemSlave{s}) operate exclusively within their designated frequency bands. By carefully designing the matching network and canceling various interference, a customized \SystemMaster communicates simultaneously with multiple \SystemSlave{s} across distinct bands. \SystemName maintains high compatibility with commercial RFID systems and \revision{presents a tag cost of less than 10 cents}. We implement a 5-band \SystemName system and evaluate its performance under various settings. The results demonstrate a fivefold increase in read rate, reaching up to 5000 reads per second.
\end{abstract}

%---------------------------------- CCS Concept ------------------------------------
%% CCS Concept Area, will be generated. The code below is generated by the tool at http://dl.acm.org/ccs.cfm. Please copy and paste the code instead of the example below.

\begin{CCSXML}
  <ccs2012>
    <concept>
      <concept_id>10010583.10010588.10011669</concept_id>
      <concept_desc>Hardware~Wireless devices</concept_desc>
      <concept_significance>500</concept_significance>
    </concept>
    <concept>
      <concept_id>10003033.10003106.10003112.10003238</concept_id>
      <concept_desc>Networks~Sensor networks</concept_desc>
      <concept_significance>500</concept_significance>
    </concept>
  </ccs2012>
\end{CCSXML}
\ccsdesc[500]{Hardware~Wireless devices}
\ccsdesc[500]{Networks~Sensor networks}

%---------------------------------- Keywords ------------------------------------
\keywords{Backscatter; RFID; FDMA; RF computing}

%----------------------------------- Content -------------------------------------
\maketitle
% !TeX root = ../main.tex

%-------------------------------------------------------------------------------
\section{Introduction}
%-------------------------------------------------------------------------------
\revision{RFID, known for its battery-free operation, presents a compelling solution for a broad range of Internet of Things (IoT) applications \cite{ahmed2023battery, RFSpy, RFIdentity}. In sectors including logistics \cite{Tagoram, RFID_Plus}, supply chain \cite{MetaSight, RF-Chord}, warehouse management \cite{dodds2023handheld, RF-DNA}, and wireless sensing \cite{See_Through_Walls, li2023go, RFWise}, the proliferation of RFID has witnessed an explosive growth in its deployment \cite{RevisitingCardinality, RevisitingMissingTag}.}

\revision{In many applications, such as industrial and logistics, real-time and high-throughput ID collection becomes crucial. For example, large-scale logistics centers handle incoming and outgoing items in bulk, often transported by forklifts. Hundreds of tagged items are concentrated within a few meters of the reader and must be scanned simultaneously to ensure seamless inventory tracking. Similarly, on high-speed production lines, such as in electronics assembly, product components arrive in batches and bursts. Within an extremely short time window, usually just a few milliseconds, dozens of tags must be read rapidly to enable real-time status updates.}

\begin{figure}[t]
    \centering
    \includegraphics[width=0.47\textwidth]{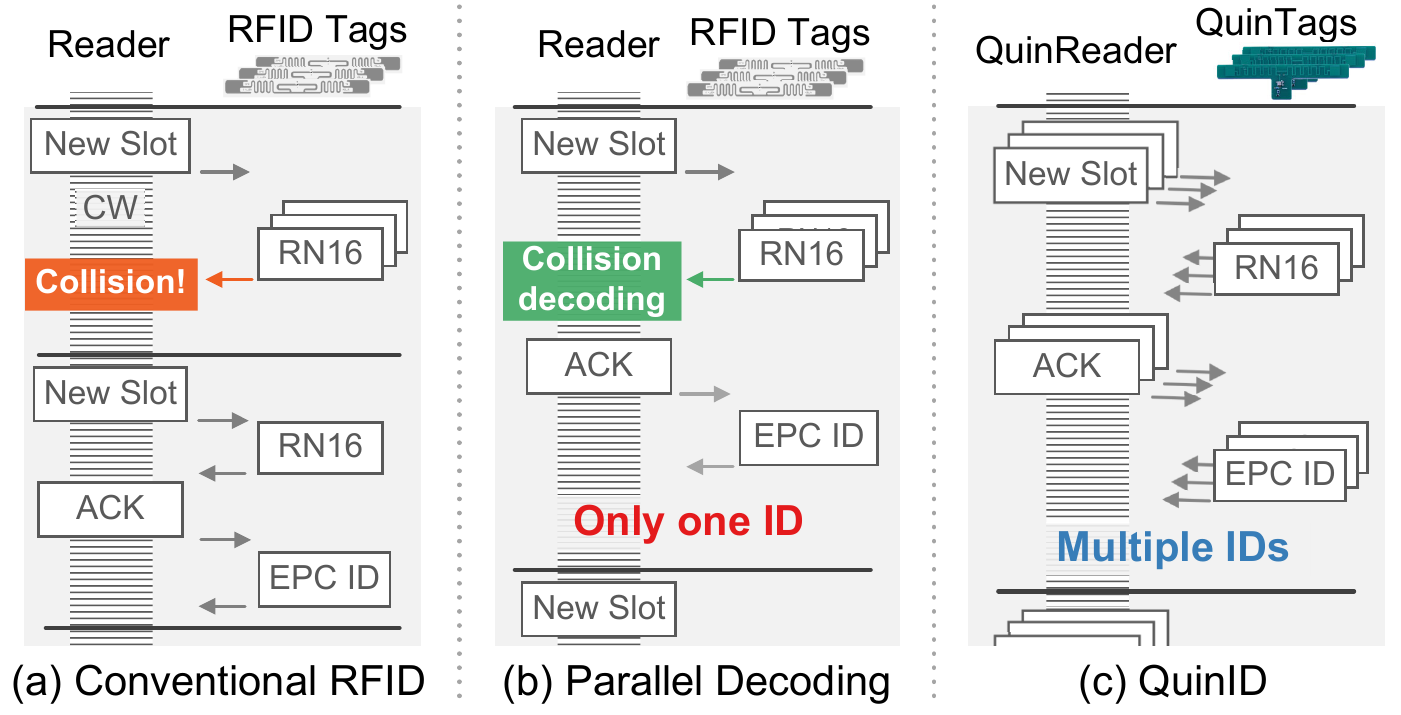}
    \vspace{-0.4cm}
    \caption{\SystemName advances from parallel transmission decoding to fully parallel reading in frequency domain.}
    \label{fig:comparison_with_existing}
    \vspace{-0.5cm}
\end{figure}

In order to enhance the efficiency of ID collection, a key research focus has been achieving parallel RFID communication. Traditional RFID employs ALOHA-based protocol, EPC Gen-2, to interrogate tags one by one \cite{EPCStandard}. To avoid collisions, \revision{the EPC standard mandates exchanging a 16-bit random number (RN16) as the handshaking process before reading the tag ID (Fig. \ref{fig:comparison_with_existing}(a)).} In the pursuit of parallelism, recent works \cite{Laissez-Faire, FlipTracer, Hubble, Physical_Layer_Collision_Recovery, Come_and_Be_Served} adopt the ``parallel decoding'' approach, involving decoding collided signals from parallel backscatter transmissions. By leveraging subtle signal features, they propose various algorithms to disentangle interleaved signals from distinct tags effectively.
\begin{figure}[t]
    \centering
    \includegraphics[width=0.45\textwidth]{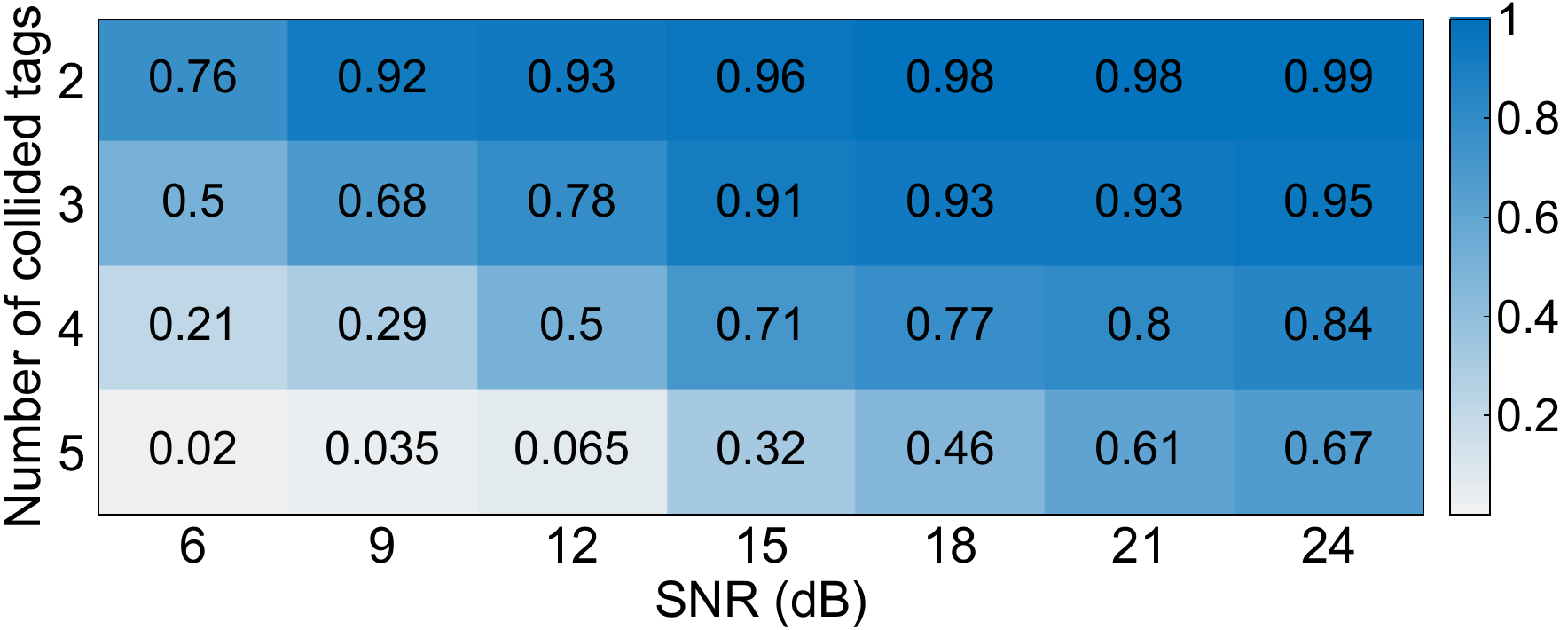}
    \vspace{-0.4cm}
    \caption{Successful decoding rate of collided RN16s using a representative algorithm from \cite{FlipTracer}.}
    \label{fig:collision_decoding_rate}
    \vspace{-0.5cm}
\end{figure}

Despite the extensive efforts, no existing approach achieves genuine and practical parallel RFID reading. When applied to RFID systems, parallel decoding algorithms \revision{only marginally accelerate the reading by decoding the collided RN16 handshaking signals mandated before transmitting the tag ID}, as shown in Fig. \ref{fig:comparison_with_existing}. This fundamental limitation still confines them to a time-division multiple access (TDMA) operation, resulting in a mere 20\% increase in the read rate in our evaluation. Further, in practical applications, these algorithms exhibit unstable performance due to varying collision numbers and fluctuating channel conditions. We implement a representative algorithm from \cite{FlipTracer} and simulate its ideal successful decoding rate, as shown in Fig. \ref{fig:collision_decoding_rate}. In real-world scenarios, collisions between tags with different signal-to-noise ratio (SNR) levels can further reduce this rate.

We observe that while tags operate across wide frequencies, readers typically only excite them within a narrow band. In this paper, we explore \textit{the potential of going beyond conventional TDMA restrictions and achieving FDMA-based fully parallel RFID communication}. The reader is expected to establish reliable communication with multiple tags simultaneously across different bands in both the uplink and downlink directions. Bringing this high-level concept into practice, however, faces serval critical challenges.

\noindent\textbullet\ \textbf{Limited energy of the tag.}
Traditional FDMA operates at the digital baseband, \revision{consuming energy far exceeding RFID tag's constrained power budget (\textasciitilde 1$\mu$W).} Recent studies \cite{DigiScatter} suggest introducing various frequency shifts on the tag to generate FDMA backsactter signal. Nonetheless, the power incurred still exceeds the affordable level of RFID tags.

\noindent\textbullet\ \textbf{High frequency selectivity requirement.}
UHF RFID operates within the 902-928MHz ISM band. A functional FDMA demands interference-free communication across channels. This requires high frequency-selectivity on the tag, especially as parallelism increases. Traditional tags, however, operate across a wide bandwidth and lack such selectivity.

\noindent\textbullet\ \textbf{Compatibility with commercial RFID.}
RFID systems have been widely deployed across various industries and applications, with billions of tags in circulation. An FDMA solution should ensure compatibility with existing systems in two aspects: 1) \revision{FDMA-enabled tags should remain readable by commercial readers, avoiding a complete overhaul of infrastructure, especially at stages where parallel reading is not required.} 2) FDMA readers should support the coexistence of FDMA-enabled and traditional tags, ensuring exclusive reading on the former when the latter is also present.
\begin{figure}
    \centering
    \includegraphics[width=0.45\textwidth]{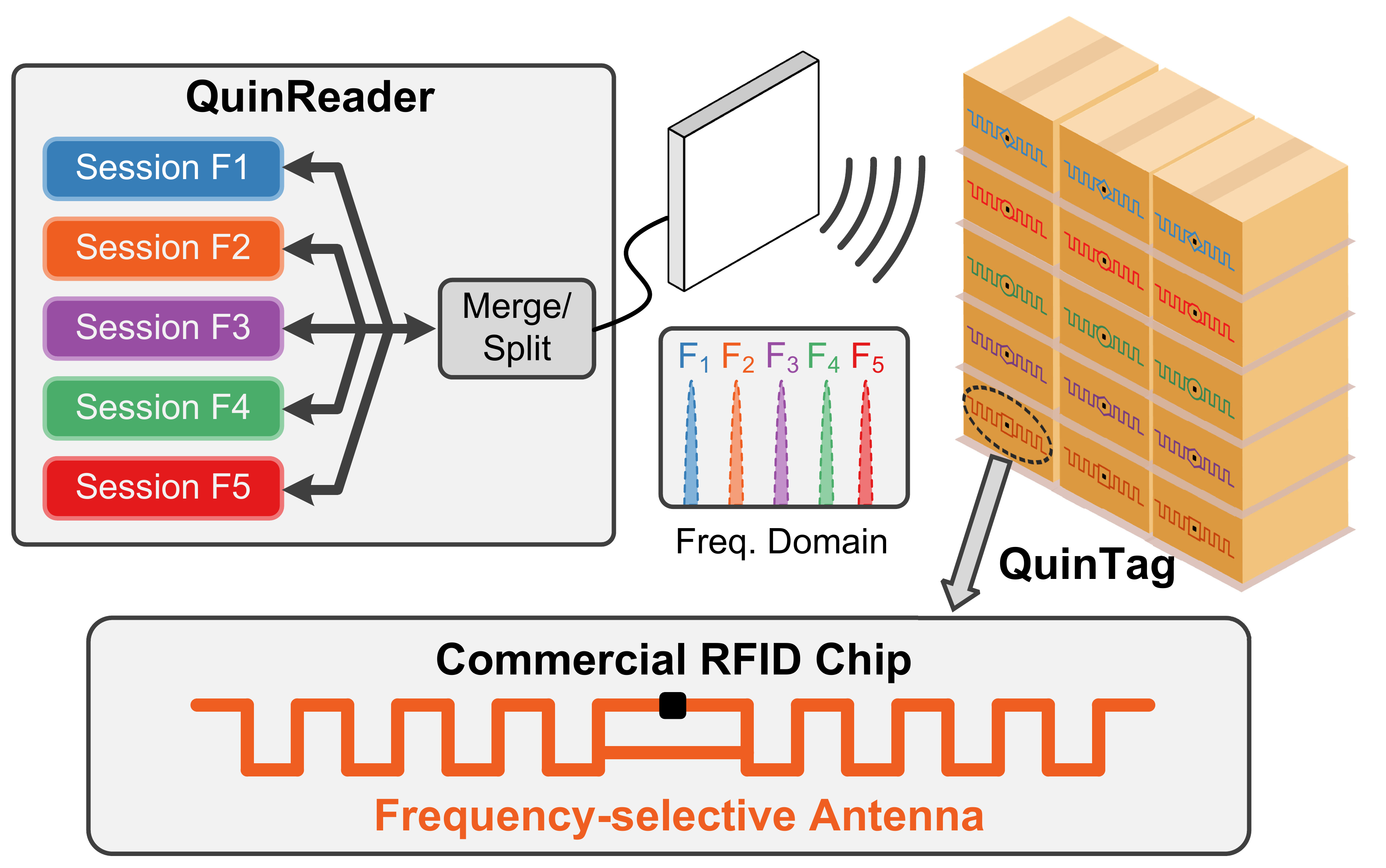}
    \vspace{-0.4cm}
    \caption{\SystemName achieves FDMA-based fully parallel RFID with frequency-selective antenna.}
    \label{fig:introduction}
    \vspace{-0.5cm}
\end{figure}

To tackle the above challenges, we present \textbf{\SystemName}, the first FDMA-based fully parallel RFID system. Instead of designing a digital FDMA circuit, we explore the frequency selectivity of the tag antenna based on its RF filtering capabilities and battery-free nature. Leveraging the frequency-selective antenna, the \SystemName tag (\SystemSlave) responds solely to excitation signal within its specific frequency band. \SystemSlave{s} are thus distributed across the whole band, each selecting a carrier from the \SystemName reader (\SystemMaster) signal, thereby enabling parallel RFID sessions in the frequency domain, as shown in Fig. \ref{fig:introduction}. The design of \SystemName requires addressing issues from both the hardware and software:

\noindent\textbullet\ \textbf{Fully passive ultra-selective filtering antenna.} 
The key to achieving \SystemSlave's functionality lies in introducing the frequency-selective antenna while maintaining a battery-free and cost-effective design. Although conventional antennas possess filtering capabilities, they cannot meet \SystemSlave's selectivity requirements. Upon a thorough exploration, we find that surface acoustic wave (SAW) filters are capable of fulfilling \SystemSlave's needs (\S\ref{subsec:tag_explore}). To obtain optimal efficiency on the filtering antenna, we carefully fine-tune a decoupled matching network on the SAW's two ports (\S\ref{sec:designing_filtenna}). Our analysis indicates that this antenna has only a slight impact on \SystemSlave's reading distance performance (\S\ref{subsec:range_reduction_compensation}). \SystemSlave maintains compatibility with conventional RFID systems, as it is essentially an RFID tag using a commercial RFID chip. \revision{Commercial readers are able to read \SystemSlave due to their frequency-hopping capabilities, which are typically enabled by default per FCC regulations \cite{FCCRegulations}.}

\noindent\textbullet\ \textbf{Interference-free multi-band reading with mutual independence.}
In \SystemName, parallel reading occurs exclusively when \SystemMaster interacts with \SystemSlave{s}. By leveraging digital up/down converters, we accurately merge and separate multi-band RFID signals (\S\ref{subsec:reader_overall}). To guarantee a robust parallel reading, we effectively eliminate two critical sources of interference. First, to avoid interference from mis-excited conventional tags, we ensure independence among \SystemMaster's reading sessions (\S\ref{subsec:interference1}). Second, to eliminate mutual interference within multi-band sessions, we incorporate a crucial pre-distortion stage to compensate for power amplifier nonlinearity (\S\ref{subsec:interference2}). Additionally, we delicately design \SystemMaster's real-time demodulation algorithm and optimize its performance on FPGA hardware (\S\ref{subsec:reader_demodulate}).

\textbf{Implementation.} \revision{We implement \SystemName in its entirety, including the commercial chip-based \SystemSlave and the FPGA-based integrated \SystemMaster, demonstrating its readiness for direct productization.} We divide the entire bandwidth into five subbands after carefully considering various factors. \revision{We open source \SystemName designs\footnote{Open source files can be found at https://github.com/wonderfulnx/QuinID.}, including the first high-performance software-defined radio (SDR) based RFID reader implementation supporting all data rates in the standard.}

This paper makes the following contributions:\\
\textbf{\textbullet} We present \SystemName, the first to enable FDMA-based fully parallel RFID, running multiple mutually independent sessions in the frequency domain.\\
\textbf{\textbullet} We design a passive ultra-selective filtering antenna to enable frequency-selective operation on RFID tags, allowing seamless integration into existing systems. A specifically designed reader achieves interference-free parallel reading.\\
\textbf{\textbullet} \SystemName achieves a $K$-fold increase in the read rate if dividing the bandwidth into $K$ subbands. Our evaluation proves a fivefold improvement, reaching up to 5000 reads per second. It achieves a 5-meter reading distance \revision{with a tag manufacturing cost of less than 10 cents}.

Paper organization: \S\ref{sec:tag_design} presents the \SystemSlave design. \S\ref{sec:reader_design} covers the \SystemMaster design. Implementation details are in \S\ref{sec:implementation}, and evaluation in \S\ref{sec:evaluation}. We review related works in \S\ref{sec:related_work} and make discussions in \S\ref{sec:discussion}. We conclude this work in \S\ref{sec:conclusion}.

% !TeX root = ../main.tex

%-------------------------------------------------------------------------------
\section{\SystemSlave Design}\label{sec:tag_design}
%-------------------------------------------------------------------------------
This section explains how \SystemSlave enables FDMA with commercial RFID chips. An RFID tag consists of a fixed IC chip and a customizable antenna. \revision{Due to its power constraints, the chip cannot distinguish carrier frequencies and lacks firmware or post-production tunability.} Since FDMA cannot be implemented on the chip, we explore the antenna's filtering capability, designing a highly selective passive filtering antenna that enables the tag to respond to only one carrier.

\subsection{Exploring Filtering Capabilities}\label{subsec:tag_explore}
\begin{figure}
    \centering
    \includegraphics[width=0.45\textwidth]{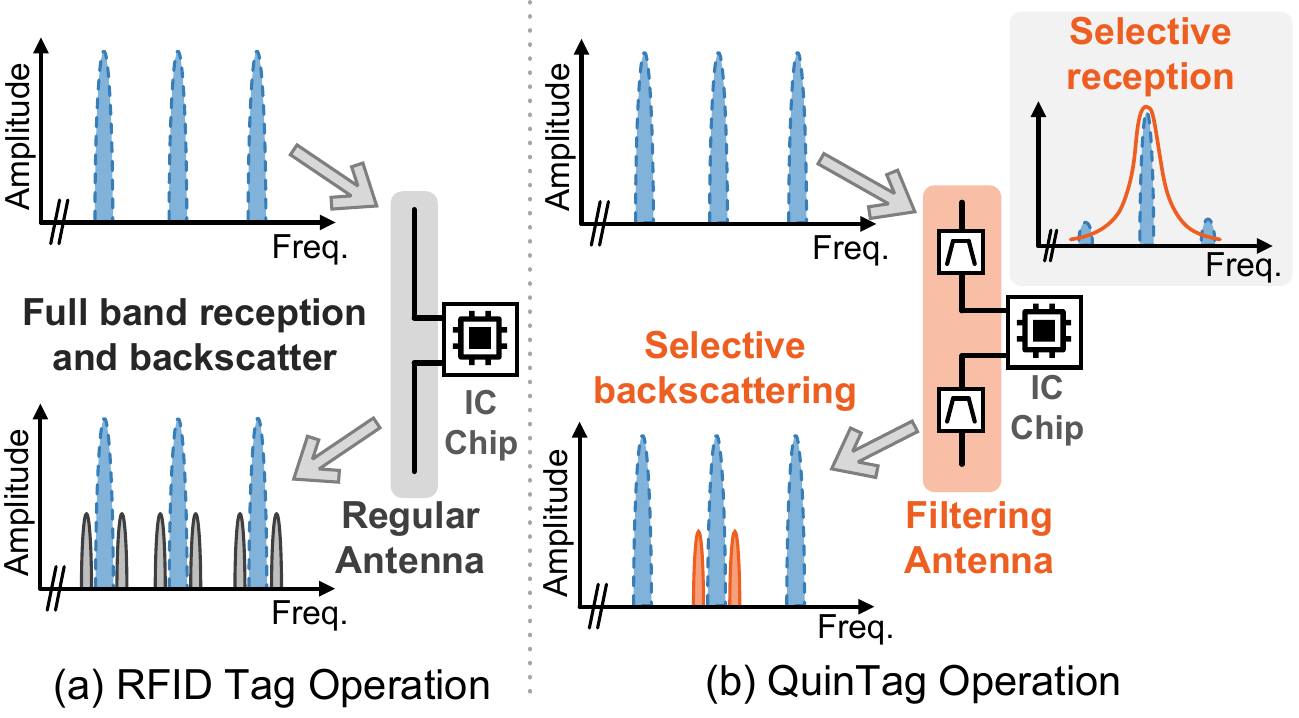}
    \vspace{-0.4cm}
    \caption{Conventional tags have wideband response, while FDMA RFID tags operate in specific bands.}
    \label{fig:quintag_operation}
    \vspace{-0.6cm}
\end{figure}

FDMA necessitates \SystemSlave to operate solely within its designated band for both transmission and reception. Unlike conventional tags that operate across the full band, \SystemSlave should be selectively excited and backscatter signals within a narrow band, as shown in Fig. \ref{fig:quintag_operation}. This requires the antenna to permit the passage of RF signals to and from the IC chip within the operating band while effectively blocking signals outside of this band, thereby requiring filtering capabilities.

Incorporating a filtering process into the antenna gives rise to a filtering antenna \cite{filtering_antenna}. Designing a suitable antenna for \SystemSlave has to satisfy the following two requirements:\\
\textbullet\ \textbf{Selectivity:} UHF RFID operates within the 902-928MHz ISM band. Dividing this band into multiple subbands imposes a substantial selectivity requirement.\\
\textbullet\ \textbf{Size and cost:} RFID applications demand lightweight and easily manufacturable tags, the same applies to its antenna.

\noindent \textbf{Challenges analysis.} The frequency selectivity of RF filtering arises from the resonant structure. Typically, the frequency selectivity of a resonant structure is quantified by its quality factor (Q-factor):
\begin{equation}\label{eq:quality_factor}
    Q=\frac{f_c}{\Delta f}=2\pi\cdot\frac{\mbox{energy stored}}{\mbox{energy dissipated per cycle}}
\end{equation}
where $f_c$ represents the resonance frequency, and $\Delta f$ signifies the passband bandwidth, corresponding to the 3dB attenuation point. A higher $Q$ implies a sharper frequency response, thereby resulting in a greater degree of parallelism. If five subbands parallelism is required, \SystemSlave necessitates a selectivity level of less than 5MHz in the 902-928MHz band, resulting in a loaded Q-factor of at least 200.

In antenna design, LC resonators or transmission line resonators are common choices. However, parasitics in LC components and heat losses in metal lines lead to irreducible dissipated energy, limiting their Q-factors to within 100 \cite{Ludwig_RFCircuitDesign}. For example, a third-order LC filter theoretically could have a passband of 902-908MHz. However, it would require a parasitic resistance of less than $10^{-12}$ Ohms, $10^{10}$ times lower than currently achievable ones. Potential high-Q solutions like waveguide resonators or dielectric resonators confine electromagnetic waves within dielectric materials, minimizing metal losses \cite{Bhartia_RF_Filter}. However, their voluminous cavity structures and complex manufacturing procedures make them unsuitable for RFID applications.
\begin{figure}
    \centering
    \includegraphics[width=0.47\textwidth]{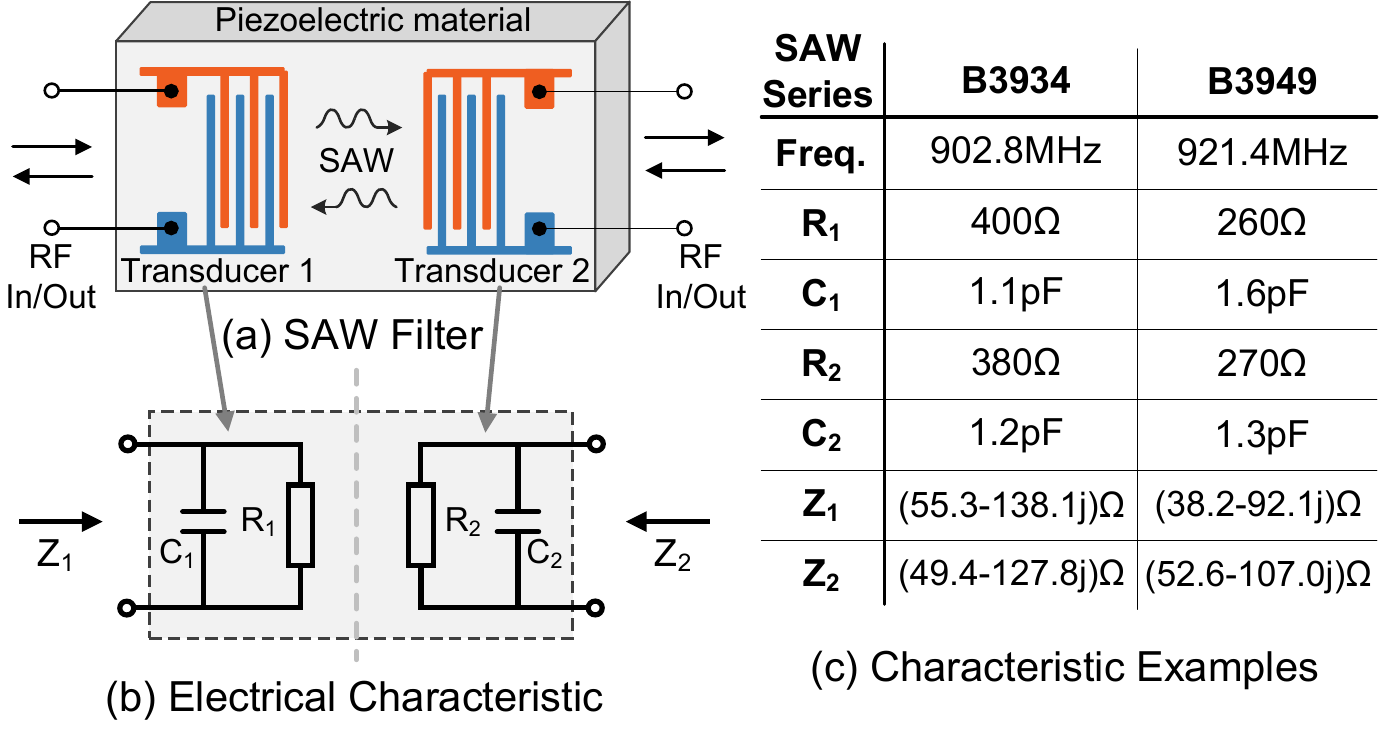}
    \vspace{-0.4cm}
    \caption{(a) SAW filter, (b) Equivalent RF circuit model, (c) Numerical model values for two series SAW filters.}
    \label{fig:SAW_filter}
    \vspace{-0.6cm}
\end{figure}

\noindent \textbf{SAW filter.}
In \SystemSlave, we exploit SAW filters to fulfill the filtering function in its antenna. A SAW filter operates by converting electrical energy into acoustic waves on piezoelectric transducers, as depicted in Fig. \ref{fig:SAW_filter}(a). Only the incidence of the RF signal near the resonant frequency induces oscillation in the transducer, resulting in an exceptionally high Q-factor of up to 1000. It establishes bidirectional filtering to support both downlink and uplink requirements. Moreover, SAW filters offer the advantage of compact size (at the IC scale) and low cost due to their straightforward manufacturing process. Generally, SAW devices costs as little as a few cents, making them a popular choice for RFID applications \cite{Turcu_2009_SAWRFID}.
\subsection{Designing SAW-based Antenna}\label{sec:designing_filtenna}
An RFID antenna needs to match its impedance with the IC chip in order to maximize the energy transmission. Specifically, the antenna impedance within the operating frequency range should be equal to the conjugate of the chip impedance. For the SAW-based frequency-selective antenna, precise matching is crucial for proper operation.

\noindent \textbf{Inconsistent impedance of the SAW filter.}
As previously mentioned, the SAW device achieves filtering through internal transducers. Each transducer consists of two interleaved metal electrodes, presenting a parallel RC characteristic in the circuit's perspective, as shown in Fig. \ref{fig:SAW_filter}(b). These two transducers remain physically unconnected, thereby generating independent capacitive impedances at the two ports ($Z_1$ and $Z_2$). Further, for SAWs working in different frequencies, the inherent differences in the transducers also yield distinct impedances. We illustrate numerical values for two SAW series in Fig. \ref{fig:SAW_filter}(c).
\begin{figure}
    \centering
    \includegraphics[width=0.45\textwidth]{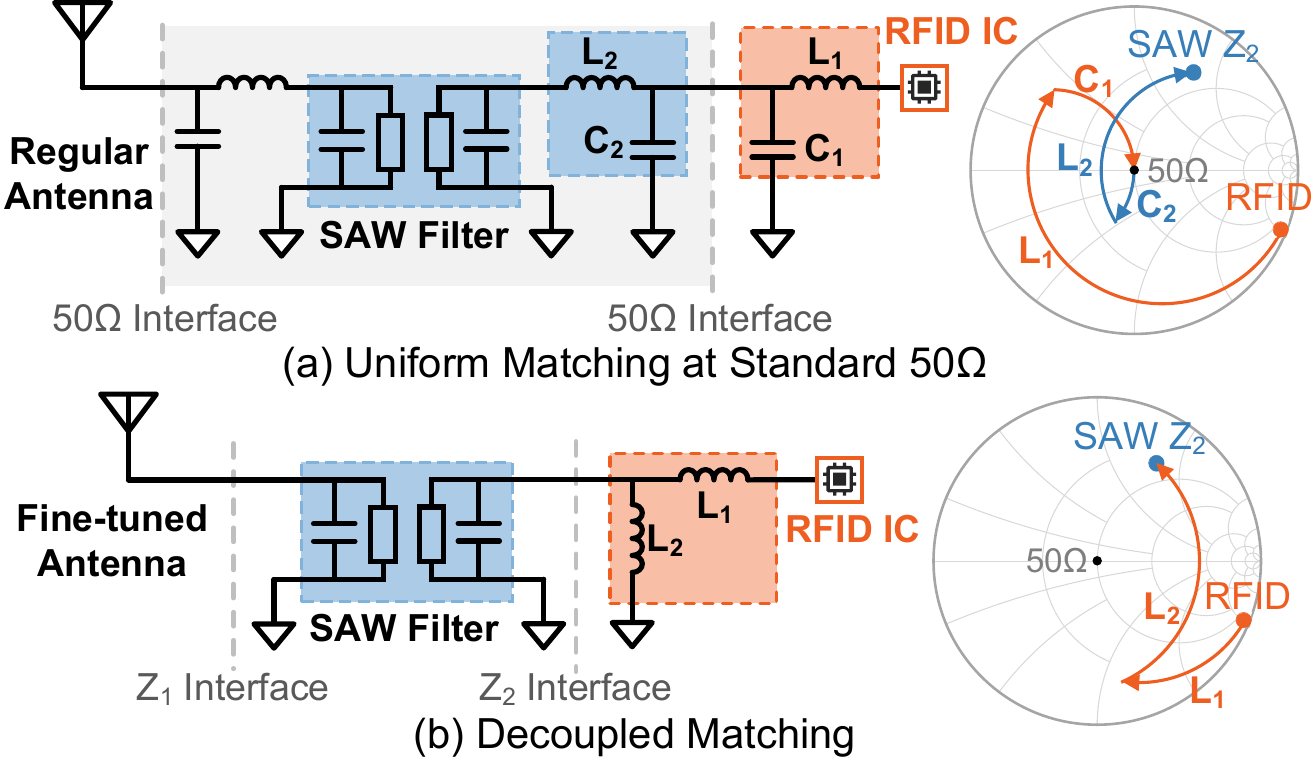}
    \vspace{-0.4cm}
    \caption{We use decoupled matching to optimize the filtering antenna's performance and reduce cost.}
    \label{fig:quintag_design}
    \vspace{-0.5cm}
\end{figure}

Considering this impedance inconsistency, a common approach to designing the SAW-based antenna would be uniformly matching both ends to a standard impedance of 50 ohms. Then, a regular antenna is attached on one end, while another matching circuit is introduced to connect with the RFID chip on the other end, as depicted in Fig. \ref{fig:quintag_design}(a). However, this conventional matching technique fails to deliver optimal performance. It necessitates using up to six matching elements for the whole tag, resulting in significant losses attributable to parasitic effects in LC components. Moreover, these additional elements substantially inflate the tag's manufacturing cost in large-scale production.
\begin{figure}
    \centering
    \includegraphics[width=0.45\textwidth]{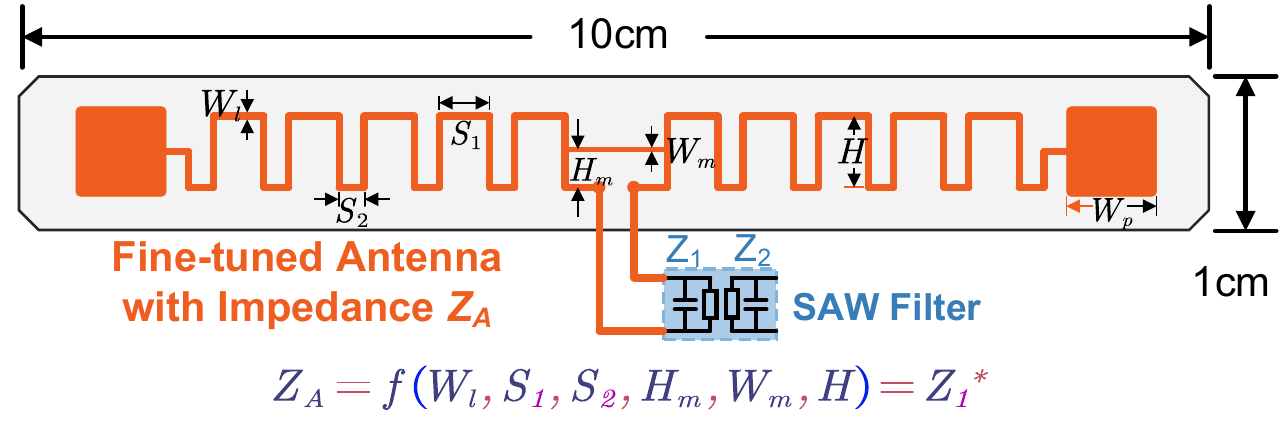}
    \vspace{-0.4cm}
    \caption{We fine-tune the meandered line's impedance to precisely match the SAW's port one.}
    \label{fig:antenna_design}
    \vspace{-0.6cm}
\end{figure}
\begin{figure*}[htp]
    \centering
    \begin{minipage}[t]{0.32\textwidth}\centering
        \includegraphics[width=0.95\textwidth]{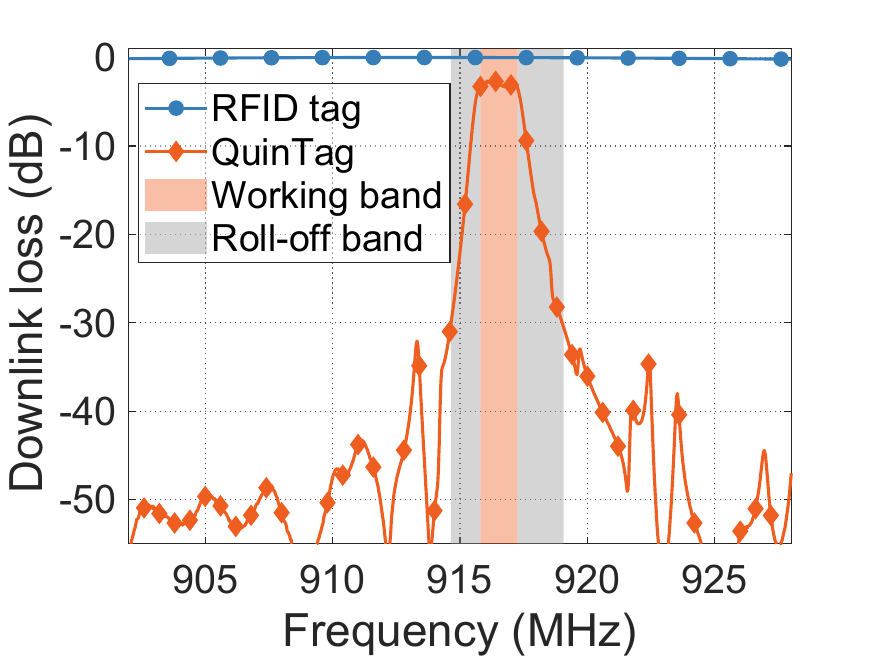}
        \vspace{-0.4cm}
        \caption{Downlink loss comparison and \SystemSlave's band division.}
        \label{fig:downlink_compare}
    \end{minipage}
    \hspace{0.15cm}
    \begin{minipage}[t]{0.32\textwidth}\centering
        \includegraphics[width=0.95\textwidth]{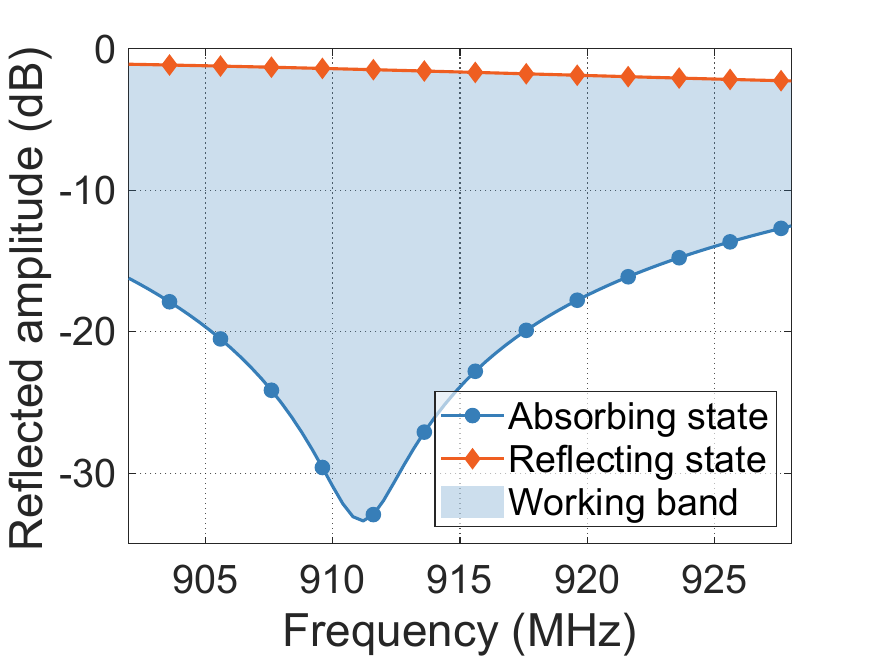}
        \vspace{-0.4cm}
        \caption{RFID tag uplink amplitude in two states (wideband backscatter).}
        \label{fig:regular_uplink}
    \end{minipage}
    \hspace{0.15cm}
    \begin{minipage}[t]{0.32\textwidth}\centering
        \includegraphics[width=0.95\textwidth]{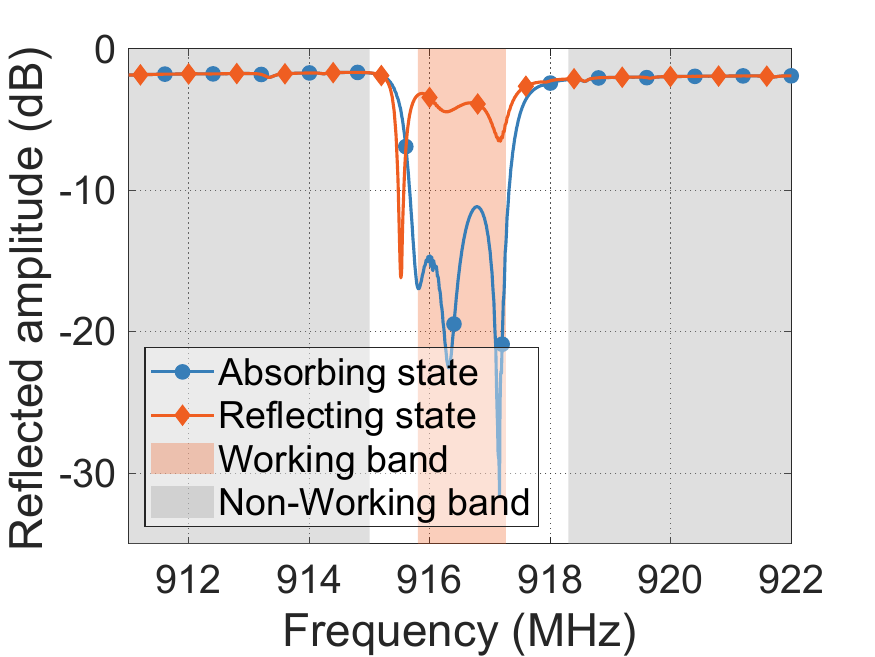}
        \vspace{-0.4cm}
        \caption{\SystemSlave uplink amplitude (narrowband backscatter).}
        \label{fig:quintag_uplink}
    \end{minipage}
    \vspace{-0.5cm}
\end{figure*}

\noindent \textbf{Decoupled impedance matching network.}
In \SystemSlave, we achieve optimal SAW-based antenna performance by directly matching the complex impedances. Analyzing the matching path from the RFID chip to the SAW on the Smith chart in Fig. \ref{fig:quintag_design}(a), we observe redundant curves. Note that the SAW's impedance point represents the conjugate of its complex impedance. This presents an opportunity to simplify the design by reducing matching elements. Instead of treating both ends as 50 ohms, we decouple them and directly match their complex impedances. We first consolidate the $C_1$, $C_2$, and $L_2$ elements into a single parallel inductor $L_2$, as demonstrated in Fig. \ref{fig:quintag_design}(b), facilitating matching between the RFID chip and the SAW. Then, the regular 50-ohm antenna element on the other end is substituted with a meandered line dipole design shown in Fig. \ref{fig:antenna_design}. It offers a 2dBi gain and maintains a similar size to typical RFID tag antennas. Through careful adjustment of its physical parameters, the impedance of this element can be finely tuned to precisely match the SAW's impedance. In this way, \SystemSlave achieves optimal matching performance using just two inductors.
\begin{figure}
    \centering
    \includegraphics[width=0.45\textwidth]{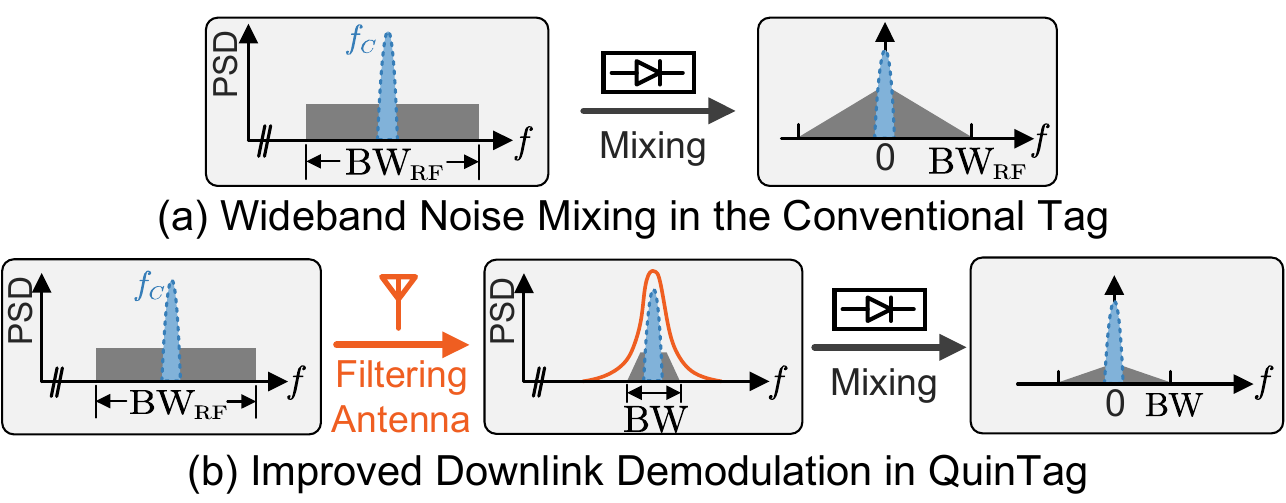}
    \vspace{-0.4cm}
    \caption{\SystemSlave filters noise to partially offset SAW insertion loss and compensate for range reduction.}
    \label{fig:noise_analysis}
    \vspace{-0.6cm}
\end{figure}

\noindent \textbf{Quick demonstration.}
For a quick proof of the above design methodology, we build and simulate \SystemSlave's frequency-selective antenna using RF simulation and full wave analysis software. A commercial Qualcomm SAW filter B3300 \cite{QualcommSAW} is used. We optimize the inductor-based matching and fine-tune the dipole so that the antenna works at 916.5MHz with a passband bandwidth of about 1.6MHz. Fig. \ref{fig:downlink_compare} compares the excitation and downlink response between this \SystemSlave and commercial RFID tags. For uplink communication involving backscatter modulation, where the tag utilizes absorbing and reflecting states for on-off keying (OOK) modulation, we present the reflective frequency response of the two states for both RFID tags (shown in Fig. \ref{fig:regular_uplink}) and \SystemSlave (shown in Fig. \ref{fig:quintag_uplink}), respectively. It can be concluded that \SystemSlave achieves frequency-selective operation in both uplink and downlink communication by utilizing the SAW-based antenna.

\noindent \revision{\textbf{Impact of material attachment.} RFID tags are often deployed on various materials like glass, wood, or acrylic, which may affect antenna impedance. However, like standard tags, \SystemSlave is minimally impacted by such attachment. Its frequency selectivity remains unaffected, as key components (SAW filter, matching network, and IC) are encapsulated and shielded from external influences. While slight frequency shifts may occur in the dipole antenna's response \cite{UHF_RFID_Antenna_Design}, its wide bandwidth (\textasciitilde 100MHz) still covers the 902-928MHz range. Thus, the tag's placement does not significantly affect its frequency selectivity or assigned sub-band.}

\noindent \textbf{Number of subband divisions.}
\revision{Dividing the RFID band into $K$ FDMA subbands enables a $K$-fold increase in reading capacity but requires interference-free operation across bands.} Although SAW filters exhibit sharp frequency responses, they are not ideal rectangles; instead, the response gradually decreases as the frequency deviates from the resonant point, forming a roll-off section. To address this, we segment the 902-928MHz band into working and roll-off bands, as depicted in Fig. \ref{fig:downlink_compare}. \SystemMaster should avoid reading tags in roll-off bands to prevent interference. \revision{Considering this constraint alongside the SAW filter's passband bandwidth and the need to maintain appropriate energy in each band,} we split the ISM band into \textbf{five} \revision{(i.e., $K$=5)} distinct FDMA subbands. Each subband is associated with a specific type of frequency-selective antenna, which is optimized and tuned using the methodology outlined above. It is also achievable to use a larger number of subband divisions, but it results in less energy and narrower data bandwidth distributed across each subband. Ultimately, this trade-off should be made according to specific application needs.

\subsection{Compensating for Range Reduction}\label{subsec:range_reduction_compensation}
While the SAW filter introduces superb frequency selectivity, it inevitably brings insertion losses of about 3dB, visible in Fig. \ref{fig:downlink_compare} and may harm \SystemSlave's reading range. Fortunately, its filtering property compensate for this reduction.

RFID systems range is determined mostly by the downlink communication range \cite{RevisitingRFID, How_Manufacturers}. As the distance reaches the threshold, the reader can still power up the tag and receive its backscatter signal. However, the tag can no longer demodulate the downlink command. This is because wideband noise is introduced alongside the signal, creating self-mixed noise on the tag's envelope detection, as depicted in Fig. \ref{fig:noise_analysis}. A calculation of the minimum detectable power ($P_m$) shows a heavy impact from the noise bandwidth \cite{Huang_Noise_Analysis}:
\begin{equation}\label{eq:noise_analysis}
    P_m = 4 \mbox{S}_m B_s K_T+2 K_T\cdot\sqrt{4B^2_s \mbox{S}_m^2 + B_n B_s\mbox{S}_m}~~,
\end{equation}
where the minimum decodable SNR $\mbox{S}_m$ and $K_T$ are constants \revision{for a specific design}, $B_s$ and $B_n$ denote the downlink signal and noise bandwidths, respectively. \revision{Although passive RFID lacks amplification, this calculation still applies. Their passive demodulators simply yield higher $K_T$ and $\mbox{S}_m$ values. Additionally, since the envelope detector compares its output to an averaged envelope \cite{Facen_RFID_Tag_FrontEnd}, the specific threshold voltage does not affect $P_m$. The same applies to the receiver's specific impedance values, as long as matching is maintained.}

The wideband nature of conventional tags brings a thermal noise spanning up to 100MHz. In contrast, \SystemSlave significantly filters noise down to approximately 2MHz, improving the downlink detectable power by about 2dB. While it cannot completely eliminate the range reduction, it ensures an acceptable communication range for RFID applications. We evaluate this compensation in \S\ref{subsubsec:quintag_on_RFID_reader}.

% !TeX root = ../main.tex

%-------------------------------------------------------------------------------
% \section{Interference-free Multi-band Interrogation}
\section{\SystemMaster Design}\label{sec:reader_design}
%-------------------------------------------------------------------------------
We now introduce how \SystemMaster effectively reads \SystemSlave{s} across multiple bands. Using digital up/down converters, it separates and merges multi-band signals. We eliminate interference from conventional tags as well as inter-band sessions. We also outline key components enabling high rate and real-time reading on FPGA platforms.

\subsection{Supporting Multi-band Transmission}\label{subsec:reader_overall}
RFID readers excites tags with a single tone and modulate it with pulse-interval encoding (PIE) for command transmission. Tags backscatter at the frequency of a few hundreds kilohertz, forming a narrow band signal around the carrier. In \SystemName, multiple such signals appear at different frequencies, each carrying its \SystemSlave's backscatter data.

\revision{A straightforward approach to reading multi-band tags is to use multiple conventional readers simultaneously, each set to a different band and equipped with corresponding RF filters. However, the cost and complexity of purchasing and deploying multiple readers scale with the number of sub-bands, making this approach impractical. It also limits usability in mobile applications, such as handheld devices. Instead, we adopt an integrated \SystemMaster design that processes multi-band signals within a single device.}
\begin{figure}
    \centering
    \includegraphics[width=0.45\textwidth]{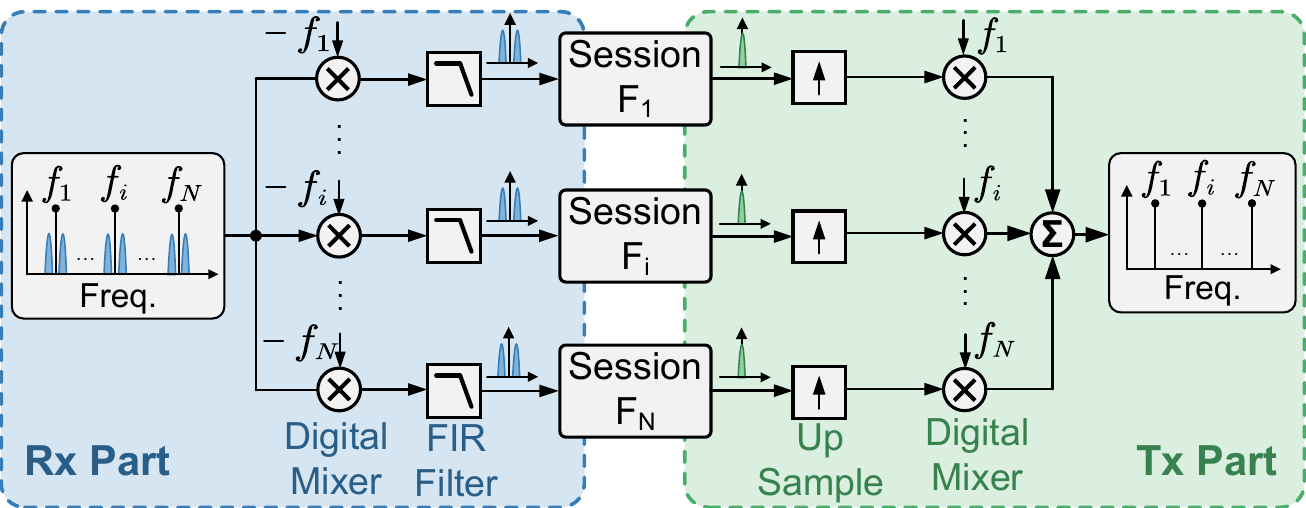}
    \vspace{-0.4cm}
    \caption{\SystemMaster uses digital up/down converters to merge and extract signals in various bands.}
    \label{fig:pre_process}
    \vspace{-0.6cm}
\end{figure}

\revision{In our integrated \SystemMaster design, we merge and separate individual RFID signals at each frequency to enable flexible multi-band processing.} As shown in Fig. \ref{fig:pre_process}, \SystemMaster employs a digital down converter (DDC) to isolate each backscatter signal. This process involves shifting the target band to the center, filtering out other bands, and feeding the resulting signal to the standard reader processing logic. A digital up converter (DUC) merges excitation signals from multiple sessions via upsampling and mixing, achieving the function opposite to DDC. To enable real-time FPGA operation, filtering and upsampling are staged to minimize resource usage and latency without sacrificing performance.

In this way, reader sessions in each band are separated and can run in parallel, reusing standard RFID reader processing. Existing collision decoding algorithms can also be seamlessly integrated to further enhance performance. However, robust multi-band reading in practice requires addressing two critical sources of interference: from conventional tags and between frequency bands.
\subsection{Avoiding Conventional Tag Interference}\label{subsec:interference1}
In practice, conventional tags inevitably present within the range of \SystemMaster, although shouldn't be considered in parallel reading. Once excited, their backscatter signals span all \SystemName bands, impacting system efficiency. Therefore, \SystemMaster needs to avoid exciting these tags.
\begin{figure}
    \centering
    \includegraphics[width=0.45\textwidth]{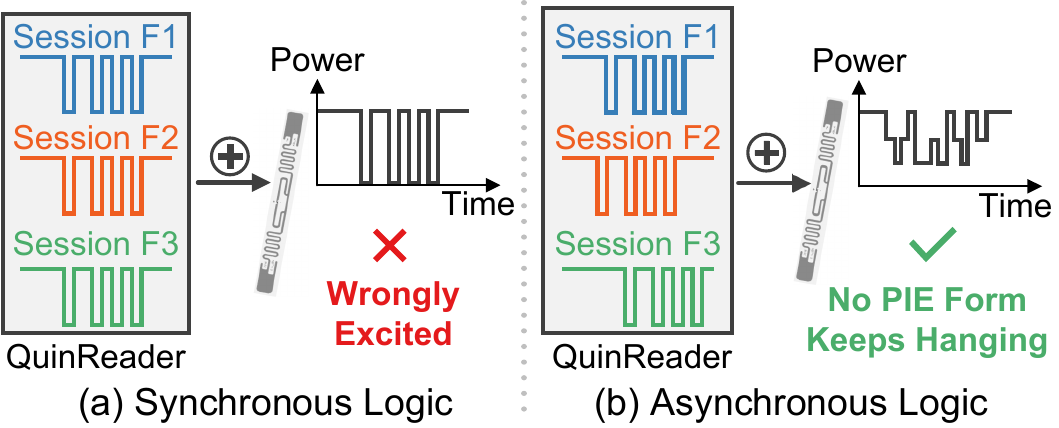}
    \vspace{-0.4cm}
    \caption{Asynchronous downlink signals effectively avoids exciting conventional tags.}
    \label{fig:async_reader}
    \vspace{-0.5cm}
\end{figure}

We find that applying time delays among multi-band downlink signals effectively prevents excitation of conventional tags. Due to their wideband nature, these tags aggregate the downlink power across bands and detect the envelope for demodulation. If \SystemMaster's downlink signals are perfectly synchronized, as depicted in Fig. \ref{fig:async_reader}(a), they can decode the command and be unintentionally excited. In contrast, the asynchronous transmission of downlink signals results in irregular variations in the aggregated envelope, shown in Fig. \ref{fig:async_reader}(b). Due to the stringent requirements on the power variation and timing of the PIE signal, the tags do not recognize this envelope as correct downlink commands. In this way, although the conventional tags may receive power from \SystemMaster, they remain dormant and do not backscatter.

To maintain the asynchronous delay consistently, we ensure that sessions across bands remain independent of each other. These sessions naturally generate asynchronous downlink signals because of the difference in command durations, tag clock offsets, and the distribution of empty slots. It's exceedingly rare for these signals to align perfectly and form a precise command structure. Even if alignment occurs momentarily, such as during the initial reading cycle, the varying durations ensure that synchronization is quickly lost. 

Nevertheless, over time, there are inevitably a few moments of synchronization as the system operates continuously. In such instances, \SystemMaster can disregard these incorrectly excited conventional tags by refraining from responding to their RN16s. This capability arises from the fact that conventional tags backscatter across all bands, resulting in multiple occurrences of the same RN16 in various sessions simultaneously. We evaluate the impact of conventional tags on \SystemMaster thoroughly in \S\ref{subsubsec:eva_conventional_tag_impact}.

\subsection{Canceling Inter-band Interference}\label{subsec:interference2}
\SystemName separates parallel RFID signals across different bands without frequency overlapping, theoretically posing no interference. However, in practice, we still find inter-band interference caused by non-ideal RF components.

\begin{figure}
    \centering
    \includegraphics[width=0.45\textwidth]{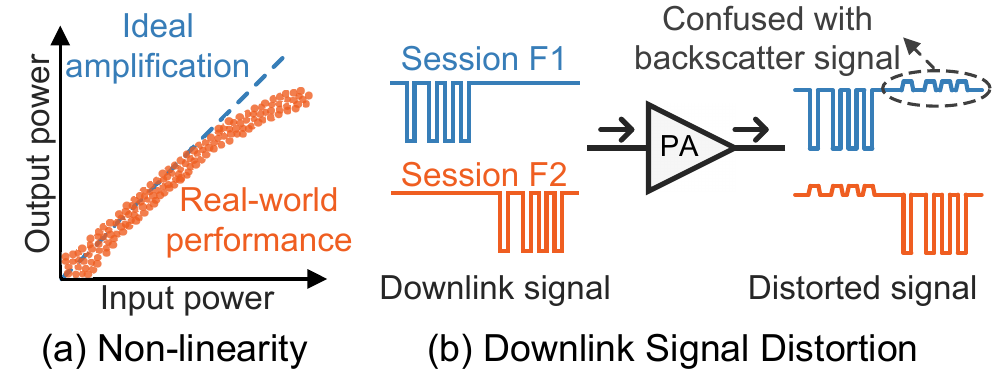}
    \vspace{-0.4cm}
    \caption{Non-linearity of the power amplifier distorts the downlink signal, causing inter-band interference.}
    \label{fig:pa_non_linear}
    \vspace{-0.5cm}
\end{figure}

Before being fed into the antenna, the multi-carrier PIE downlink signal needs to be amplified (e.g. to 27dBm). Ideally, the power amplifier (PA) linearly amplifies the input signal by a factor $A$. For an input signal $x(t)$, the output signal $y(t)$ should be $y(t)=A\cdot x(t)$. However, real-world amplifiers are constructed with nonlinear active components like transistors, resulting in inherent nonlinearity such as gain saturation during operation \cite{NonlinearityPA}. As depicted in Fig. \ref{fig:pa_non_linear}(a), this leads to a decrease in actual amplification gain as input power increases. Consequently, interference among downlink signals located in various bands is observed.
\begin{figure}
    \centering
    \includegraphics[width=0.47\textwidth]{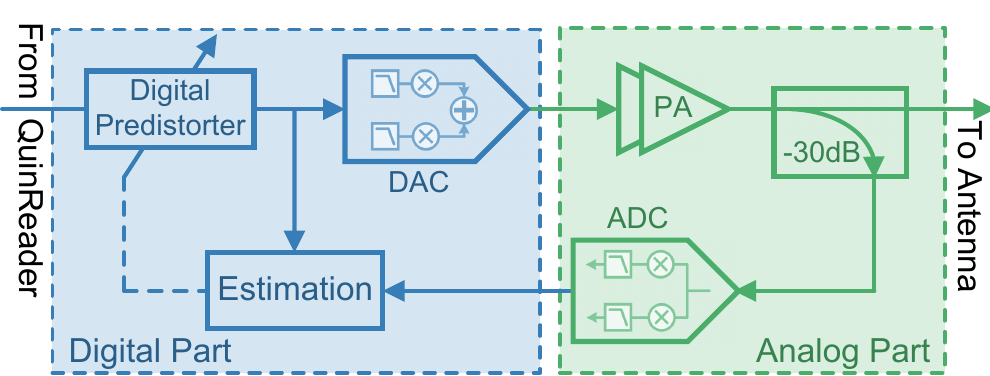}
    \vspace{-0.4cm}
    \caption{\SystemName utilizes DPD before multi-carrier downlink transmission to linearize the amplifier.}
    \label{fig:digital_pre_distortion}
    \vspace{-0.5cm}
\end{figure}

Fig. \ref{fig:pa_non_linear}(b) shows an example. Initially, when both sessions transmit a carrier, the amplified signals remain stable. At this point, the total input power reaches its maximum, causing the amplifier to become saturated. Subsequently, when one session stops transmitting to indicate 'LOW' in PIE modulation, the input power is halved. However, this reduction in input power does not proportionately decrease the output power. The actual output power will be higher than half of its original. Consequently, we observe an increase in the power transmitted by the other session. This interference closely resembles the tag's backscatter signal and may confuse the receiver or even collide with the tag signal.

To mitigate such inter-band interference, one may avoid the saturation region by restricting the transmit power. Alternatively, one may employ a mapping-based approach, which entails fitting the nonlinear curve to a function $f$ and computing its inverse function $f^{-1}$. Linearity can be restored by adjusting the actual feed power to $f^{-1}(x)$ for each desired input power $x$. However, both techniques fall short of addressing the interference effectively. This is due to the presence of the memory effect in the PA, whereby the output at a given moment depends not only on the current input but also on previous input values. Consequently, the PA's nonlinearity depicted in Fig. \ref{fig:pa_non_linear}(a) manifests as scattered points rather than a smooth curve.

\begin{figure*}[htp]
    \centering
    \begin{minipage}[t]{0.63\textwidth}\centering
    \includegraphics[width=0.98\textwidth]{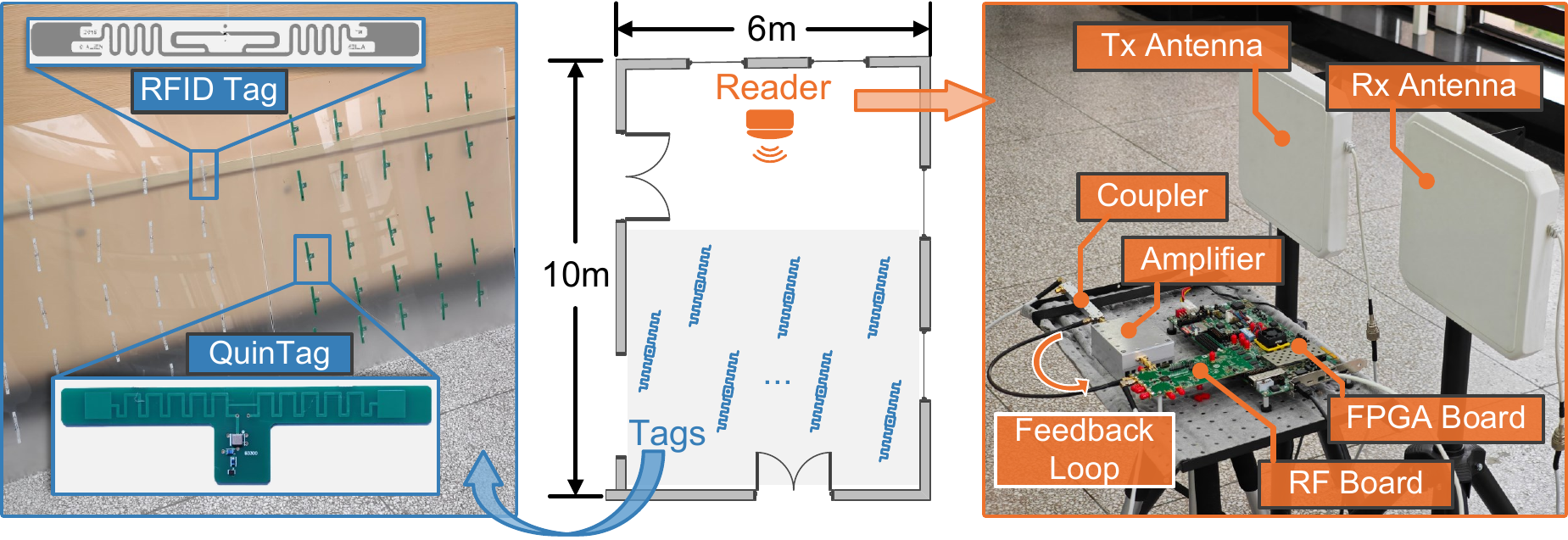}
    \vspace{-0.4cm}
    \caption{\SystemName's implementation and experiment deployment.}\label{fig:implementation_deploy}
    \end{minipage}
    \hspace{0.3cm}
    \begin{minipage}[t]{0.3\textwidth}\centering
    \includegraphics[width=0.98\textwidth]{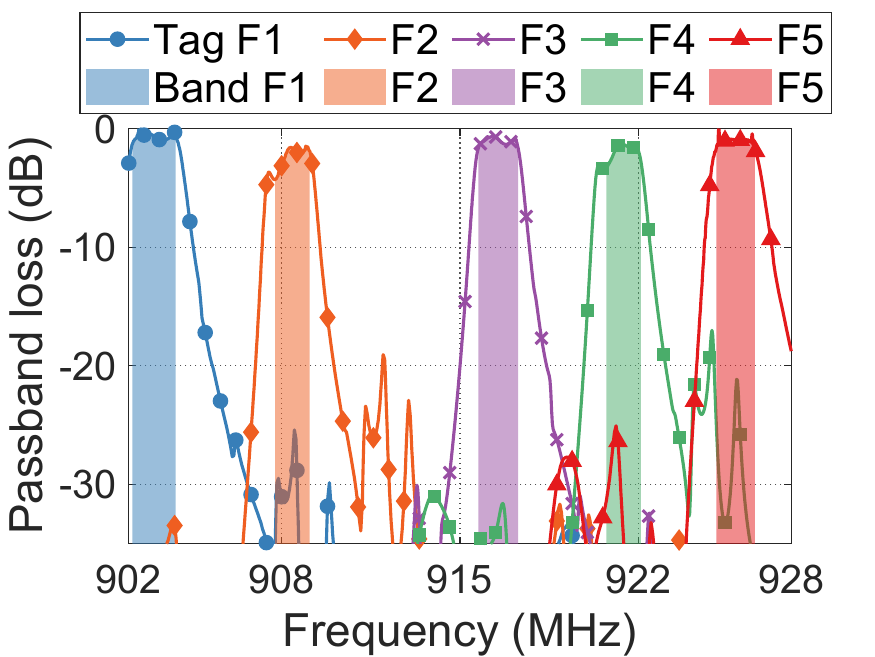}
    \vspace{-0.4cm}
    \caption{\SystemSlave{s}' filtering performance and working bands.}\label{exp:band_division}
    \end{minipage}
    \vspace{-0.6cm}
\end{figure*}

To fully eliminate inter-band interference, we use Volterra series \cite{GMP_Model} to characterize the amplifier's nonlinearity:
\begin{equation}\label{eq:volterra_series}
    y(n) = \sum_{k=0}^{K-1}\sum_{m=0}^{M-1} a_{km} x(n-m) |x(n-m)|^k
\end{equation}
At any moment, the amplifier's output $y(n)$ is determined by the present and past inputs $x(n-m)$, along with a coefficient matrix $a_{km}$, which is to be estimated for a specific amplifier.

We introduce digital pre-distortion (DPD) to linearize the power amplifier, as illustrated in Fig. \ref{fig:digital_pre_distortion}. A portion of the amplified signal is split using a directional coupler and fed back to the digital processing section. By analyzing both the pre-amplified and post-amplified digital signals, an estimator evaluates the nonlinear model and calculates $a_{km}$. Subsequently, a DPD unit pre-distorts the original transmission signal, utilizing the inverse function technique similar to that mentioned above, but applied to the IQ signal instead of just the power. This approach ensures that the final transmitted signal achieves linear amplification. We integrate DPD into \SystemMaster using the ADI ADRV9375 RF board \cite{ADRV9375} and evaluate its performance in \S\ref{subsubsec:dpd_evalute}.

\subsection{FPGA-based Real-time Demodulation}\label{subsec:reader_demodulate}
\revision{To demonstrate \SystemMaster's readiness for productization, we design a real-time RFID demodulation process optimized for FPGA. This FPGA-based design is also essential for supporting all EPC-specified backscatter link frequencies (BLFs) and enabling DPD, both of which are unattainable with common SDRs like USRP. In USRP-based readers, inherent delays in data streaming between the USRP and its host machine (caused by UDP transmission and kernel scheduling) hinder timely ACK command responses as required by the standard, limiting the BLF to as low as 40kHz \cite{USRP_Latency}.}

Upon receiving the backscatter signal, we first remove the DC component induced by carrier excitation. Time synchronization is then achieved through matching the packet preamble, which also aids in channel estimation and equalization. To accommodate the RFID chip's high and fluctuating clock drift, a symbol synchronizer is incorporated to perform clock recovery. Finally, the backscatter bits are extracted.

The latency bottleneck in this process mainly lies in channel equalization and clock recovery. Equalization requires complex operations like division and square root, which are costly on FPGA platforms. We optimize the computation time by pre-storing results in lookup tables, replacing in-place computation with fast lookups. For clock recovery, we use a phased lock loop (PLL)-based design with a simple yet efficient Gardner timing error detector for fast and accurate drift estimation \cite{2008digital}. We evaluate this latency in \S\ref{subsec:demodulation_latency}.

% !TeX root = ../main.tex

%-------------------------------------------------------------------------------
\section{Implementation}\label{sec:implementation}
%-------------------------------------------------------------------------------
% We implement the \SystemName system capable of running \textit{five} parallel RFID sessions in the 902-928MHz ISM band, including \SystemSlave and \SystemMaster.
\subsection{\SystemSlave}
Five distinct types of \SystemSlave{s} corresponding to five bands are individually designed and manufactured on two-layer printed circuit boards (PCBs), each measuring 100\textit{mm} by 20\textit{mm}, as Illustrated in Fig. \ref{fig:implementation_deploy}. They are implemented battery-free, eliminating any need for external energy sources. Each type of \SystemSlave's implementation contains four parts:\\
\textbf{\textbullet}\ \textbf{RFID IC.} To ensure compatibility with commercial RFID, we use the NXP UCODE 7 chip \cite{UCode7}. We opt for the packaged version rather than the bare-die for mounting on the PCB.\\
\textbf{\textbullet}\ \textbf{SAW filter.} We use five Qualcomm SAW filters (B3934, B3943, B3300, B3949, and B3944 series) \cite{QualcommSAW}, with each \SystemSlave incorporating one to determine its FDMA band. The excitation (or downlink) losses for each type are shown in Fig. \ref{exp:band_division}. \revision{In each band, \SystemSlave suppresses cross-band interference by 30dB. Note that each FDMA band spans 2-3 RFID channels, allowing commercial readers to read \SystemSlave{s} within these channels.}\\
\textbf{\textbullet}\ \textbf{Fine-tuned antenna.} We design the 2dBi meandered line dipole antenna for each \SystemSlave type using CST Studio Suite. Their impedances are optimized to match the corresponding SAW's complex impedance. The antenna can also be implemented in a flexible manner for general applicability.\\
\textbf{\textbullet}\ \textbf{RFID IC and SAW matching.} We use PathWave ADS to simulate the inductor-based matching. Inductors are from muRata, and the specific values are optimized with ADS.

\subsection{\SystemMaster}
We implement \SystemMaster on an FPGA-based SDR platform, featuring a Xilinx ZC706 motherboard \cite{ZC706} and an ADRV9375 RF daughter board \cite{ADRV9375}, as shown in the right side of Fig. \ref{fig:implementation_deploy}. \revision{In addition to the streaming delays discussed in \S\ref{subsec:reader_demodulate}, processing delays introduced by tools like GNURadio for USRPs further increase latency to 400-700$\mu$s \cite{GNURadio_Latency}, limiting their BLF to 40kHz, as shown in Table \ref{tb:latency}.}

\SystemMaster includes five independent RFID reading sessions, each performing the full EPC querying process. \revision{It equally distributes power across these bands.} We implement all processing algorithms on the FPGA using Mathworks Simulink HDL Coder for minimal latency and real-time operation. A host computer connects to the FPGA board via Ethernet to interface with \SystemMaster and monitor the received IQ signal and the EPC ID.
\begin{figure*}[htp]
    \centering
    \begin{minipage}[t]{0.3\textwidth}\centering
    \includegraphics[width=0.98\textwidth]{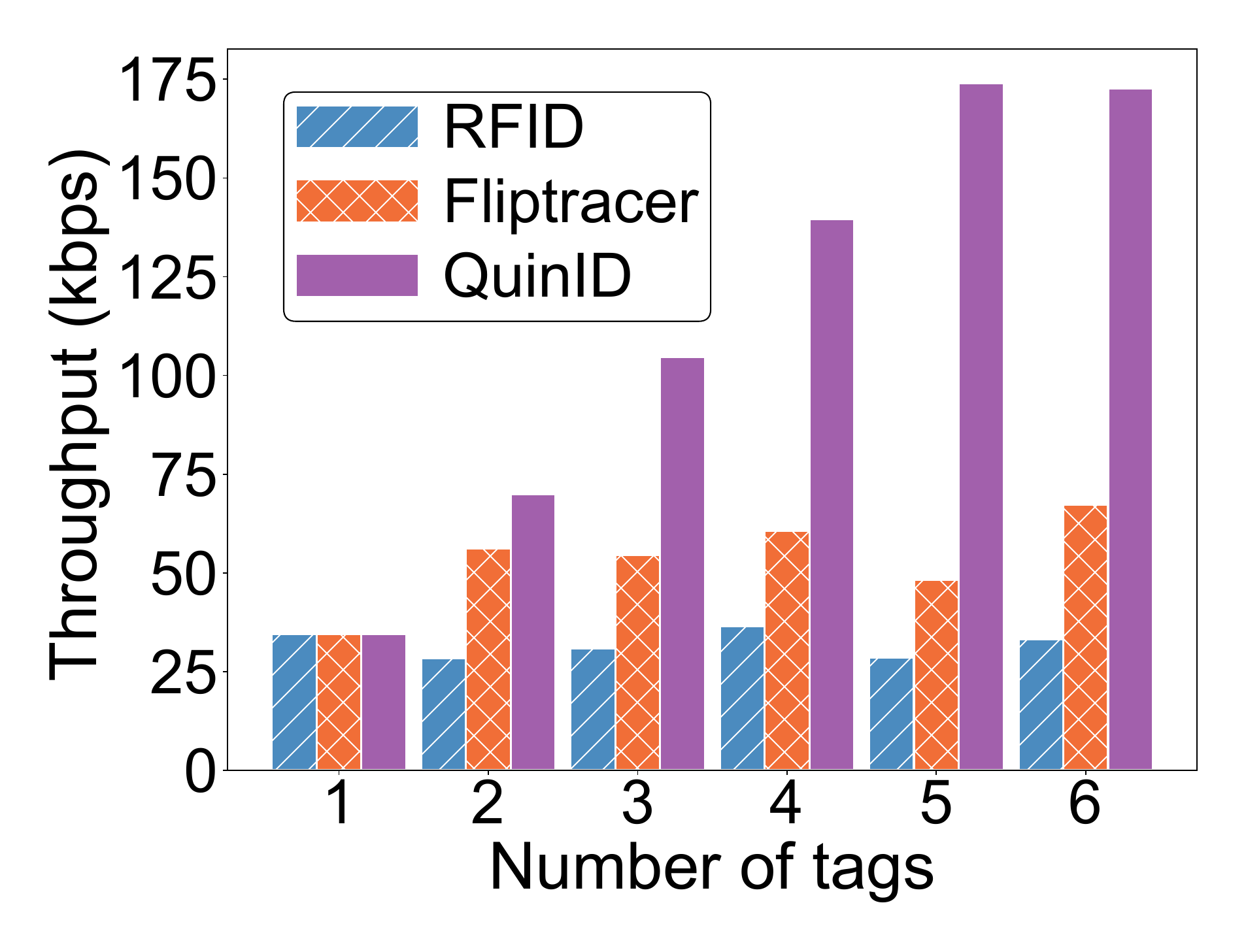}
    \vspace{-0.6cm}
    \caption{Throughput comparison with different number of tags.}\label{exp:overall_tagnum_throughput}
    \end{minipage}
    \hspace{0.3cm}
    \begin{minipage}[t]{0.3\textwidth}\centering
    \includegraphics[width=0.98\textwidth]{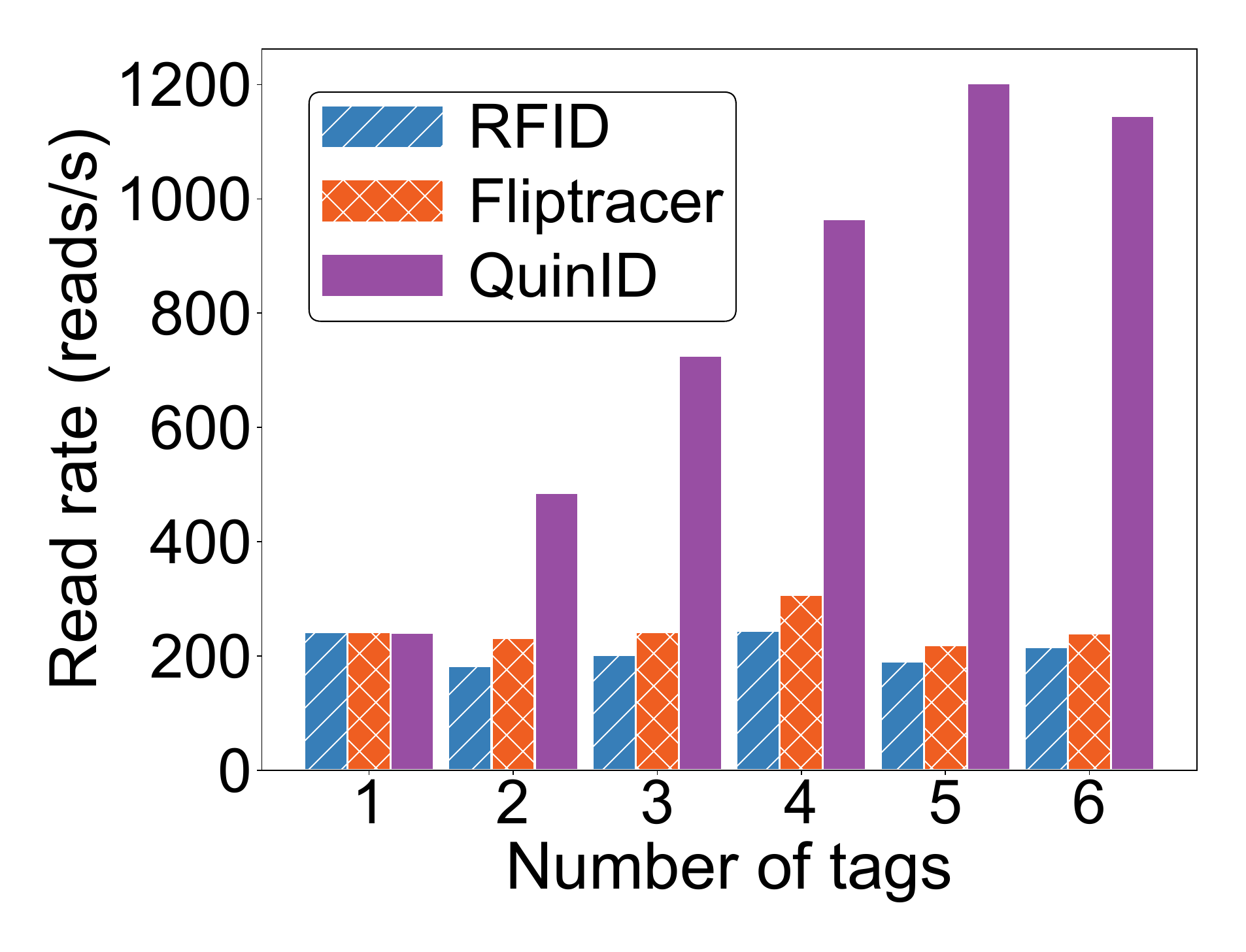}
    \vspace{-0.6cm}
    \caption{Read rate comparison with different number of tags.}\label{exp:overall_tagnum_readrate}
    \end{minipage}
    \hspace{0.3cm}
    \begin{minipage}[t]{0.3\textwidth}\centering
    \includegraphics[width=0.98\textwidth]{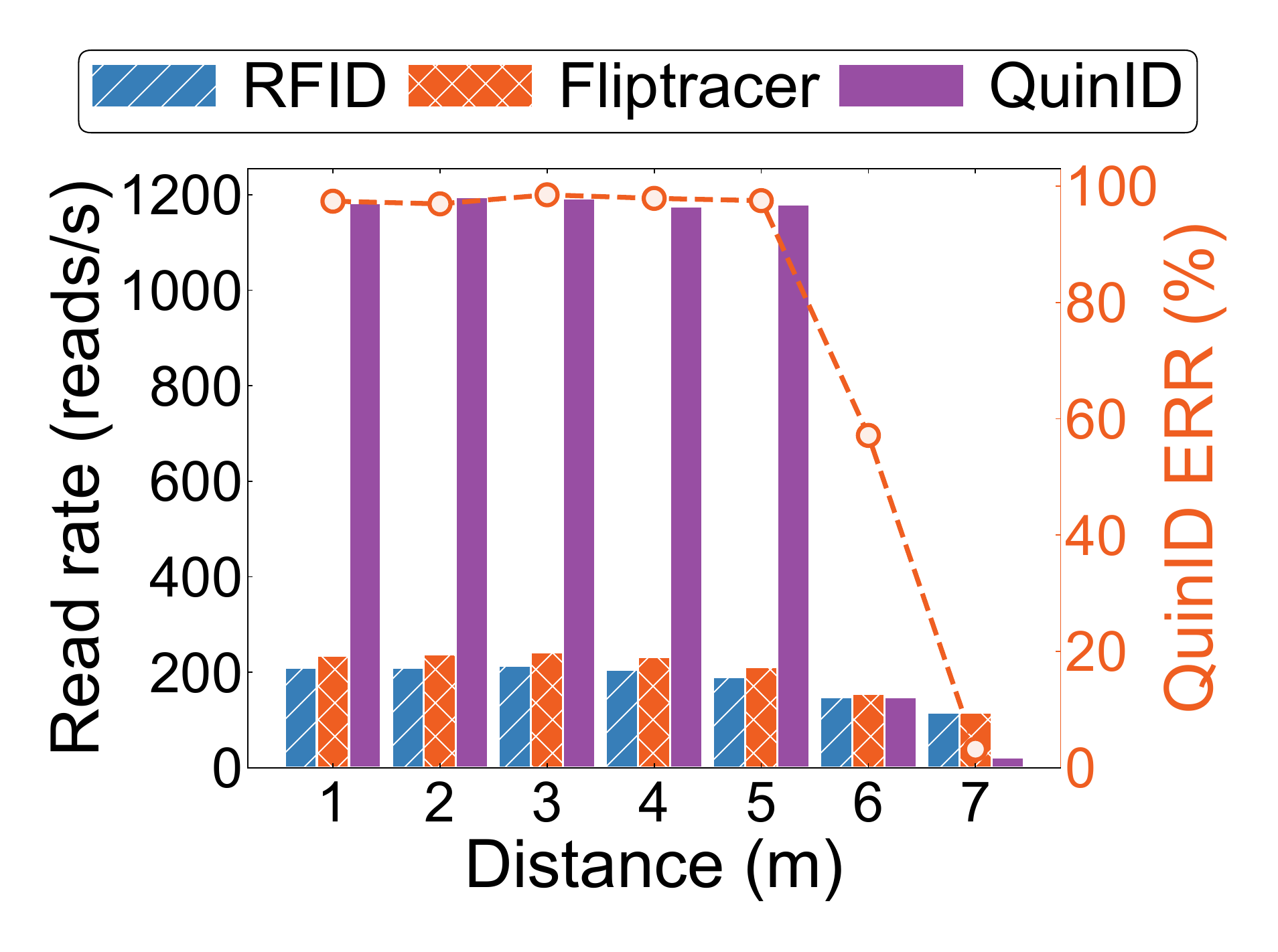}
    \vspace{-0.6cm}
    \caption{Read rate and ERR in different distances (5 tags present).}\label{exp:overall_distance_readrate}
    \end{minipage}
    \vspace{-0.5cm}
\end{figure*}

% RF and DPD part
The transmit signal from the ADRV9375 chip gets amplified to 27dBm and goes through a directional coupler. This coupler feeds a small portion of the amplified signal back to the RF board for the digital pre-distorter to cancel the inter-band interference. The signal is then transmitted to a 9dBi Laird RFID antenna, forming a 36dBm EIRP as required by FCC \cite{EPCRules}. The receiving antenna is another Laird RFID antenna connected directly to the RF board. \revision{Currently, \SystemMaster cancels RFID self-interference digitally within FPGA processing rather than using analog circuits.}

\subsection{Compatibility with Commercial RFID}
\revision{\SystemName remains compatible with commercial RFID systems, avoiding a complete infrastructure overhaul. \SystemMaster is deployed only when parallel reading of \SystemSlave{s} is needed at specific stages of an item's lifecycle. In other scenarios, commercial readers can seamlessly read \SystemSlave{s} due to their frequency hopping capabilities.}

% !TeX root = ../main.tex

%-------------------------------------------------------------------------------
\section{Evaluation}\label{sec:evaluation}
%-------------------------------------------------------------------------------
We first describe the methodology (\S\ref{subsec:methodology}) and overall performance (\S\ref{subsec:overall_performance}), then compare demodulation latency (\S\ref{subsec:demodulation_latency}) and evaluate the impact of practical factors (\S\ref{subsec:practical_impact}). Cross-band reading results follow in \S\ref{subsec:eva_cross_band}, with ablation studies in \S\ref{subsec:ablation_study}.

\subsection{Methodology}\label{subsec:methodology}
\revision{A key design goal of \SystemName is compatibility with commercial RFID. Therefore, we consider parallel decoding RFID the state-of-the-art and compare the performance with a representative scheme, Fliptracer \cite{FlipTracer}. Fliptracer accelerates reading by demodulating collided RN16s, reducing MAC layer resource waste. We also include a conventional reader as a baseline.} The performance metrics include throughput, read rate, and EPC reception ratio. \textbf{Throughput} is calculated based on all uplink packets from the tag, including both the RN16s and the EPC IDs. Since the EPC ID contains a CRC inside the packet, we calculate the \textbf{read rate} by measuring the number of CRC-passed EPC IDs per second. \textbf{EPC reception ratio (ERR)} measures the proportion of CRC-passed EPC packets to the total number of received packets.

\subsection{Overall Performance}\label{subsec:overall_performance}
We first evaluate the overall performance of \SystemName with varying numbers of tags and different distances. As shown in Fig. \ref{fig:implementation_deploy}, experiments are conducted in a hall, with the readers positioned at one end and tags placed at varying distances. To ensure a fair comparison, we implement both the standard RFID reader and the Fliptracer-based collision decoding reader using the same hardware as \SystemMaster. Both readers operate with commercial Alien ALN-9640 tags. Fliptracer's sampling rate varies with the tag data rate to maintain a fixed 10 samples per symbol for effective parallel decoding. For \SystemName, we use \SystemMaster to read \SystemSlave{s}. All readers transmit at 36dBm EIRP according to the FCC's requirements. We set the default tag bit rate as 80kbps with FM0 encoding and default distance as 2m. Since the EPC protocol requires the number of ALOHA time slots in each query round to be set to $2^Q$, we carefully select $Q$ to make the slot number close to the tag number, ensuring maximum efficiency. We let the reader operate for 10 seconds and record the packets backscattered by the tags. Unless otherwise posted, the following experiments use the same settings.

\textbf{Number of tags.} Fig. \ref{exp:overall_tagnum_throughput} and Fig. \ref{exp:overall_tagnum_readrate} compare the aggregated uplink throughput and read rate achieved with different numbers of tags, respectively. We can see that although Fliptracer increases the throughput by a maximum of $2\times$ by decoding many collided RN16s, it only improves the read rate by less than 20\% due to the TDMA nature of retrieving the tag ID. \SystemName, on the other hand, enables fully parallel reading and significantly outperforms existing schemes, achieving a maximum of $5\times$ read rate improvement. This improvement is because the \SystemName implementation operates in five distinct frequency bands. When the tag number reaches 6, the performance no longer improves since multiple tags operate in the same band. Note that the fluctuation of traditional RFID's performance comes from the varying $Q$ value. It also slightly impacts FlipTracer's improvements.

\textbf{Distance.} We compare the read rate in different reader-tag distances with 5 tags present. As shown in Fig. \ref{exp:overall_distance_readrate}, the read rate of conventional RFID decreases when the distance increases due to degradation of downlink energy. The improvement of Fliptracer also decreases as more decoding errors of collided RN16s occur. In comparison, \SystemName provides a stable and reliable $5\times$ rate improvement in 5 meters. The rate decreases significantly afterward mainly because the energy is spread into five bands.
\begin{figure*}[htp]
    \centering
    \begin{minipage}[t]{0.3\textwidth}\centering
    \includegraphics[width=0.98\textwidth]{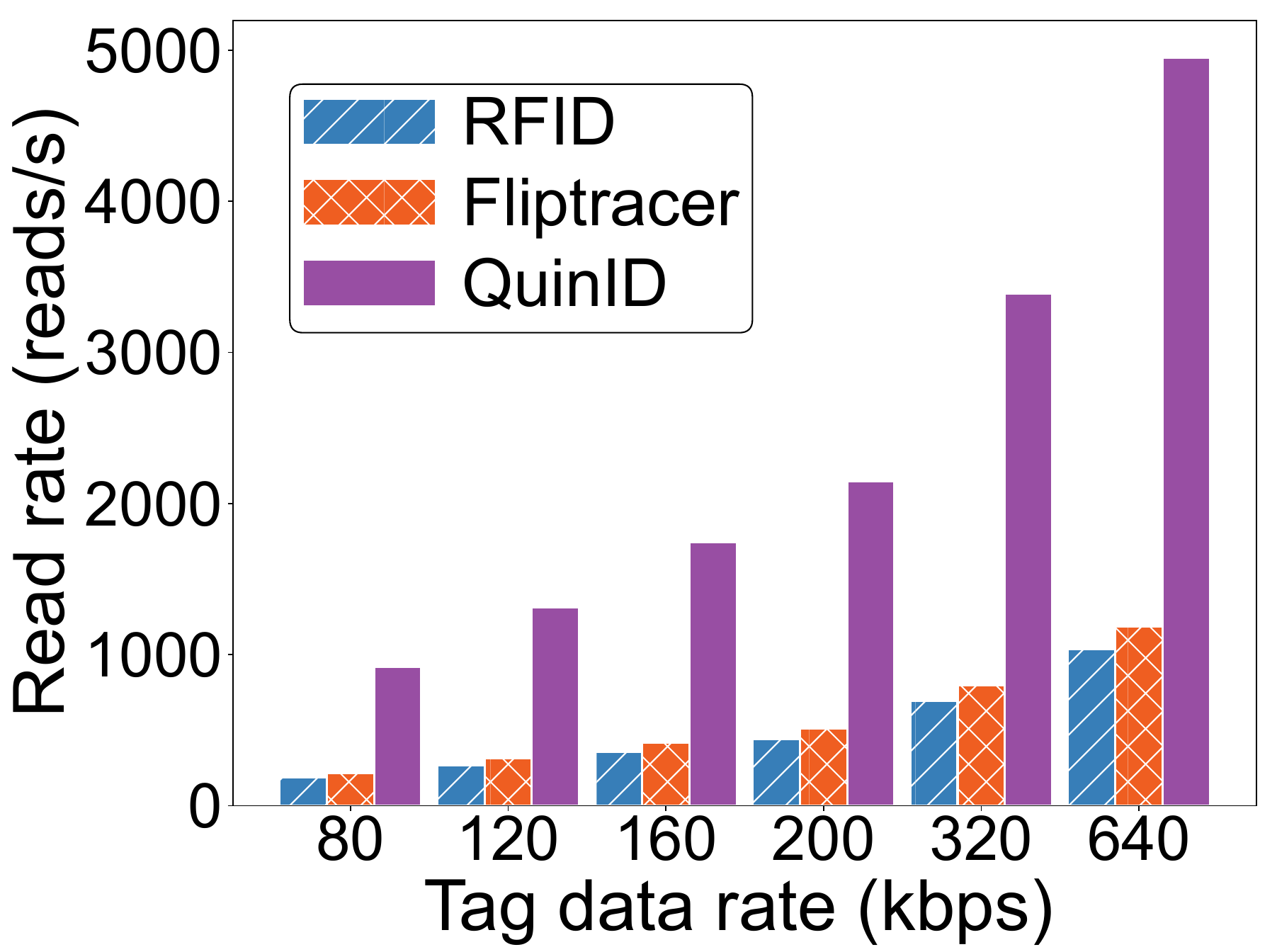}
    \vspace{-0.4cm}
    \caption{Read rate under different tag bit rates (25 tags present).}\label{exp:datarate_overall}
    \end{minipage}
    \hspace{0.5cm}
    \begin{minipage}[t]{0.3\textwidth}\centering
    \includegraphics[width=0.98\textwidth]{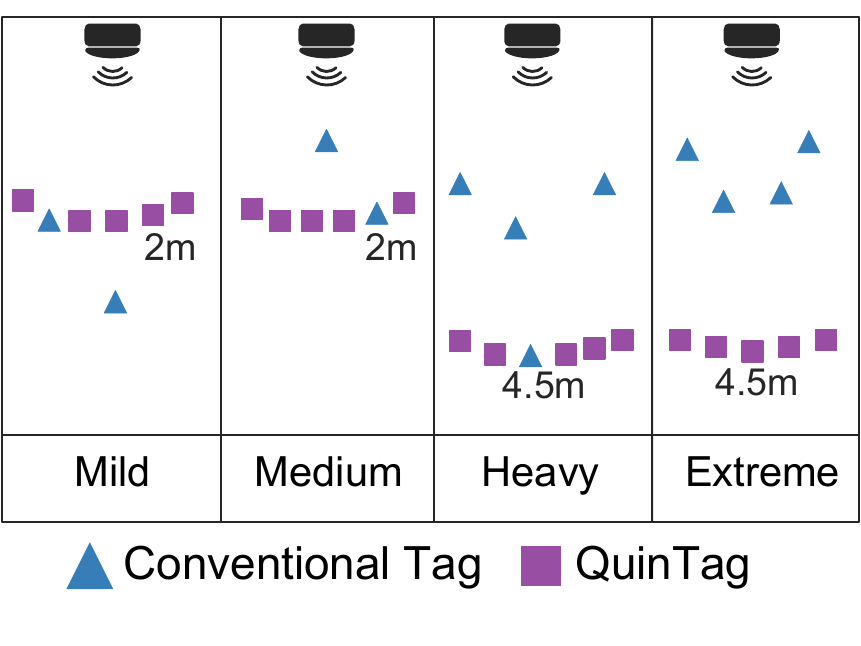}
    \vspace{-0.4cm}
    \caption{Experiment setup of conventional tags' impact.}\label{exp:conventional_tag_impact_scenario}
    \end{minipage}
    \hspace{0.5cm}
    \begin{minipage}[t]{0.31\textwidth}\centering
    \includegraphics[width=0.98\textwidth]{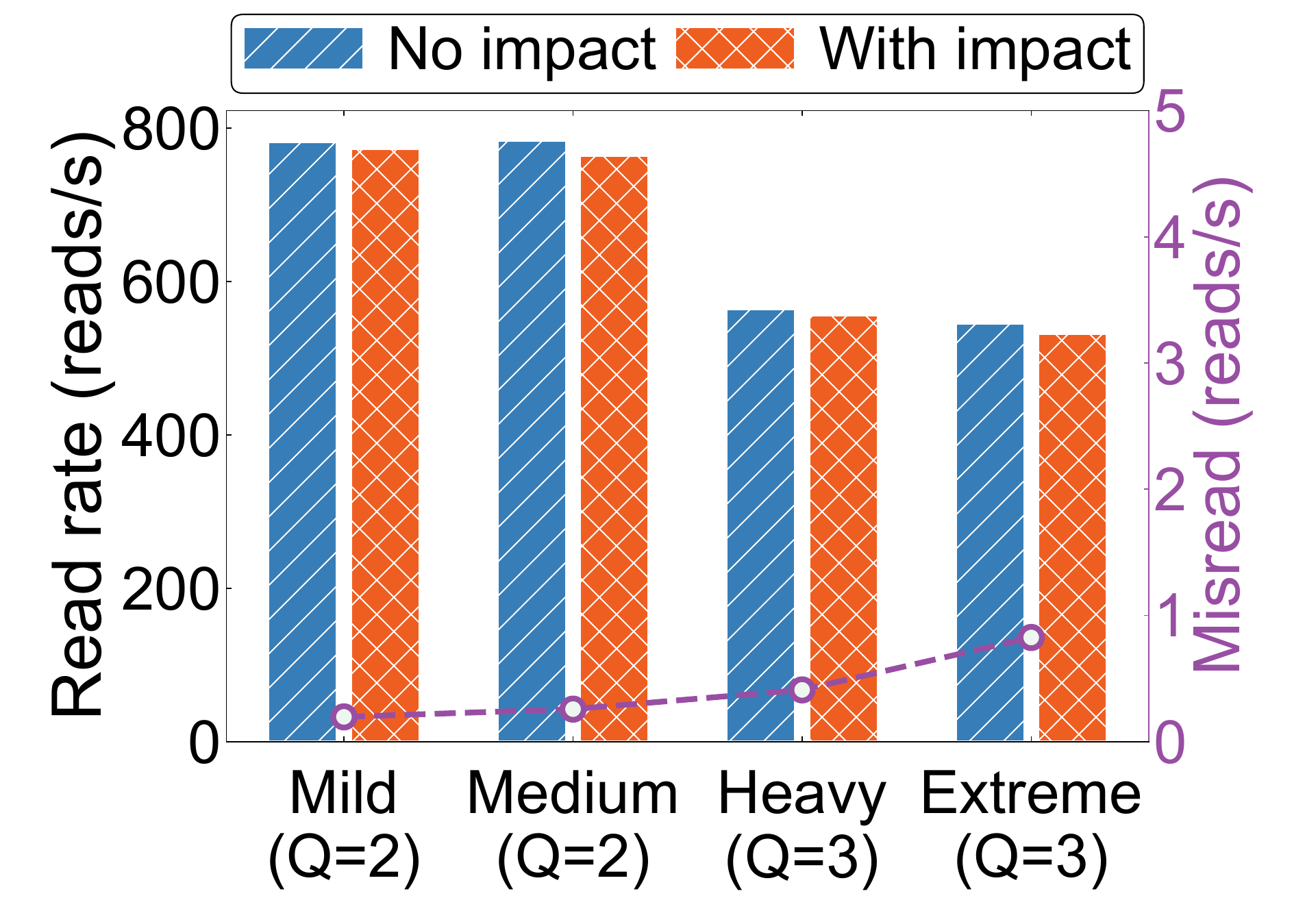}
    \vspace{-0.4cm}
    \caption{Conventional tags' impact on \SystemName.}\label{exp:conventional_tag_impact}
    \end{minipage}
    \vspace{-0.4cm}
\end{figure*}

To further investigate the bottleneck of this reading range, we show the EPC reception ratio in different distances in Fig. \ref{exp:overall_distance_readrate}. We can see that as the distance increases, more uplink CRC errors appear. It suggests that the bottleneck of \SystemName's communication range is the uplink, i.e., the reader can still wake up the tag and transmit downlink information, but the low backscatter signal SNR limits the communication distance. Therefore, \revision{one way to improve the reading range is to enhance receiver sensitivity using analog carrier cancellation circuits, as seen in modern RFID readers \cite{CarrierCancellation, CMOS_RFID_Reader}. Results from \S\ref{subsubsec:quintag_on_RFID_reader} and \S\ref{subsubsec:eva_parallelism} demonstrate the range improvement these circuits can provide. Commercial readers with analog cancellation achieve a 6-meter reading range for \SystemSlave{s} at 30dBm EIRP, while \SystemMaster requires 33dBm EIRP to achieve the same range, considering power distribution.}

\revision{\textbf{Material attachment.} In this experiment, \SystemSlave{s} are placed on cardboard to simulate real-world deployments. We also attach them to acrylic materials (as shown in Fig. \ref{fig:implementation_deploy}) in the following experiments. Both setups exhibit no observable performance difference or degradation compared to unattached cases, supporting our conclusion in \S\ref{sec:designing_filtenna}.}

\subsection{Demodulation Latency}\label{subsec:demodulation_latency}
The EPC protocol imposes strict demodulation latency requirements, mandating completion within 20 tag-bit durations. We compare the average RN16 decoding latency at different tag rates to this limit in Table \ref{tb:latency}. In this experiment, FlipTracer decodes collided RN16s in the time domain on a desktop with 13th-gen Core i7 and 64GB RAM, while \SystemName performs parallel decoding in the frequency domain on the ZC706 FPGA platform.
\begin{table}[t]
    \centering
    \vspace{0.2cm}
    \resizebox{0.48\textwidth}{!}{
        \begin{tabular}{cccccc}
        \toprule
        \textbf{Tag rate (bps)}  & \textbf{40k} & \textbf{80k} & \textbf{160k} & \textbf{320k} & \textbf{640k} \\ \hline
        \begin{tabular}[c]{@{}c@{}}\textbf{Maximum}\\ \textbf{allowed}\end{tabular} & 500$\mu$s      & 250$\mu$s       & 125$\mu$s       & 62.5$\mu$s      & 31.25$\mu$s     \\ \hline
        \textbf{FlipTracer} &
          \multicolumn{5}{c}{\begin{tabular}[c]{@{}c@{}}50.6$\mu$s (2 tags)\space\space\space\space\space\space\space97.2$\mu$s (3 tags)\\ 189.0$\mu$s (4 tags)\space\space\space\space\space275.9$\mu$s (5 tags)\end{tabular}} \\ \hline
        \textbf{\SystemName}            & 64.5$\mu$s     & 34.8$\mu$s      & 19.2$\mu$s      & 12.2$\mu$s      & 22.8$\mu$s      \\ \bottomrule
        \end{tabular}
    }
    % \vspace{0.1cm}
    \caption{Comparison of demodulation latency.}
    \label{tb:latency}
    \vspace{-1.0cm}
\end{table}

We can see that FlipTracer's latency depends on the number of collided tags but remains consistent across data rates. The former is evident, since an increase in the tag number leads to an exponential rise in IQ clusters, significantly increasing decoding complexity. On the other hand, the consistency arises because FlipTracer requires a fixed number of samples per symbol for state transition determination, regardless of the data rate. However, as the data rate increases, the protocol demands progressively shorter decoding latencies, casuing FlipTracer to miss protocol deadlines and fail to send ACK commands. In contrast, \SystemMaster's demodulation runs in parallel and ensures timely decoding, meeting the protocol requirements across all data rates.

\subsection{Practical Scenarios}\label{subsec:practical_impact}
To understand the performance of \SystemName in practical scenarios, including using higher bit rates and with the presence of regular RFID tags, we conduct the following experiments.
\subsubsection{Tag bit rate}
In practical RFID applications, a high bit rate is often required to achieve maximum reading speed. \SystemSlave achieves frequency-selective operation by leveraging the SAW-based filtering antenna, having a limited bandwidth. The other side of the coin is that this limited bandwidth can affect the bit rate of the tag since a higher bit rate requires wider bandwidth for the tag's OOK modulation. This experiment investigates \SystemName's supported bit rate.

We deploy 25 conventional RFID tags and 25 \SystemSlave{s} on an acrylic board, as shown in Fig. \ref{fig:implementation_deploy}. We vary the bit rate of all readers from 80kbps to 640kbps (the highest defined by the EPC protocol) and measure the read rate. Since FlipTracer cannot meet the latency requirements, we record the IQ signals from the standard reader for offline decoding and derive its data rate. Fig. \ref{exp:datarate_overall} shows that \SystemName maintains a stable $5\times$ improvement regardless of the tag's rate. Fliptracer, however, provides less than 20\% improvements and suffers from time-consuming decoding, especially at high rates. The result also shows that \SystemName can provide a read rate of up to 5000 per second using the highest bit rate. However, it comes with a cost associated with the tolerance of the carrier frequency offset. We evaluate this further in \S\ref{subsubsec:eva_cfo}.
\begin{figure*}[htp]
    \centering
    \begin{minipage}[t]{0.31\textwidth}\centering
    \includegraphics[width=0.98\textwidth]{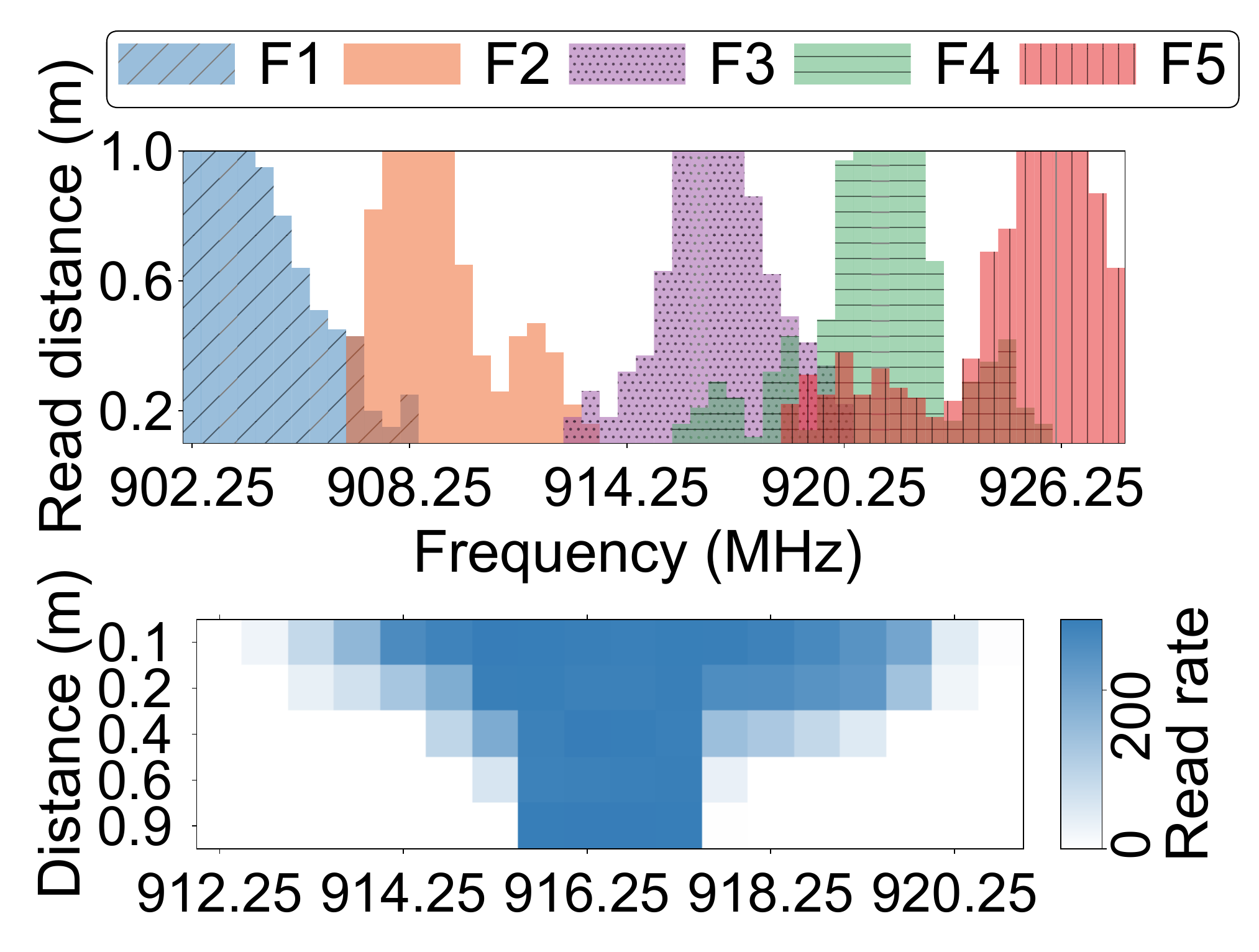}
    \vspace{-0.4cm}
    \caption{Cross-band misreading. (up): readable distance across frequencies, (down): details of band 3.}\label{exp:cross_frequency}
    \end{minipage}
    \hspace{0.3cm}
    \begin{minipage}[t]{0.31\textwidth}\centering
    \includegraphics[width=0.98\textwidth]{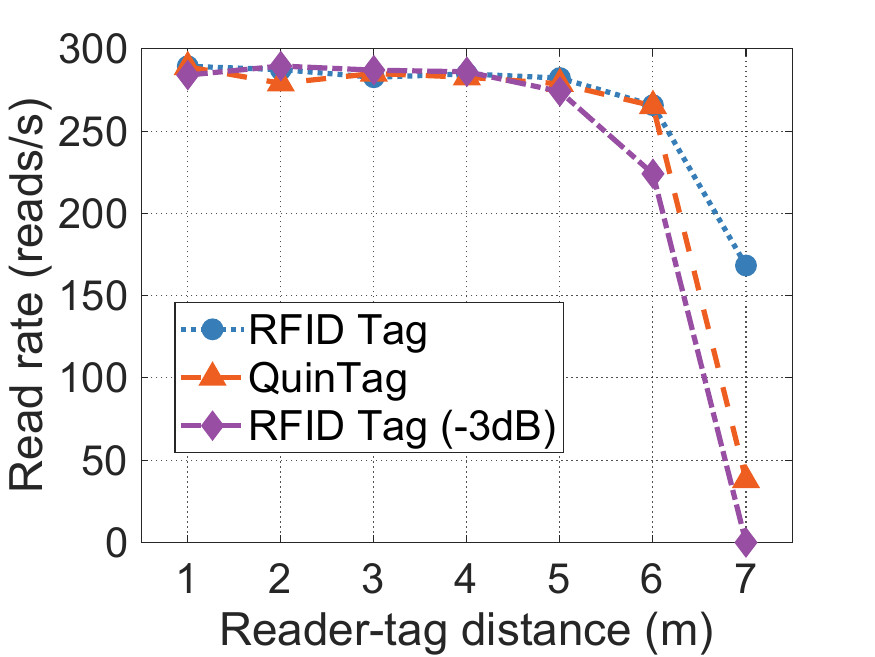}
    \vspace{-0.4cm}
    \caption{Performance comparison between \SystemSlave{s} and commercial RFID tags in commercial readers.}\label{exp:conventional_reader}
    \end{minipage}
    \hspace{0.3cm}
    \begin{minipage}[t]{0.31\textwidth}\centering
    \includegraphics[width=0.98\textwidth]{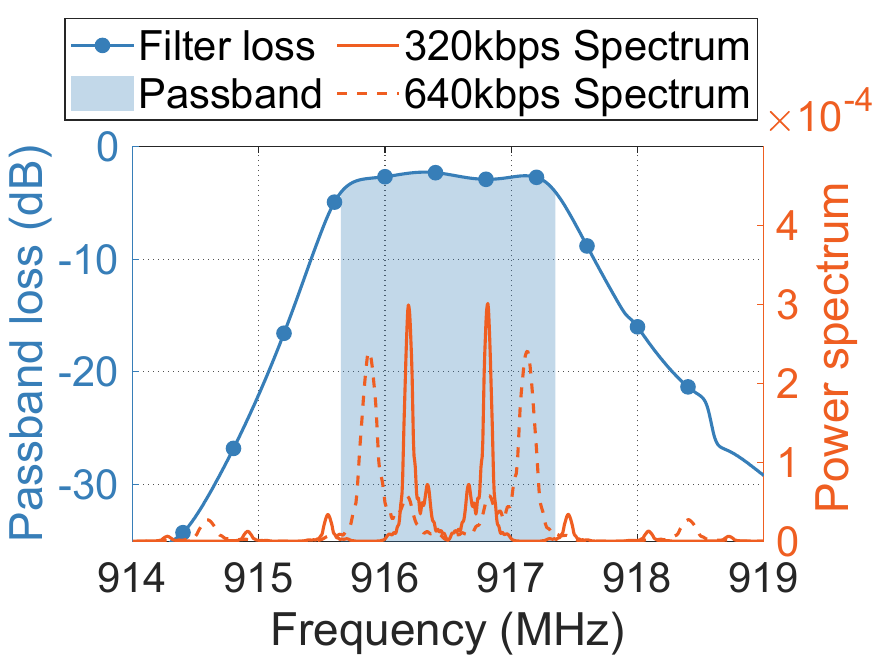}
    \vspace{-0.4cm}
    \caption{\SystemSlave supports all RFID bit rates since the spectrum falls inside its operating bandwidth.}\label{exp:datarate_spectrum}
    \end{minipage}
    \vspace{-0.5cm}
\end{figure*}

\subsubsection{Presence of conventional tags}\label{subsubsec:eva_conventional_tag_impact}
In the practical applications of \SystemName, conventional RFID tags' existence is inevitable. We set four scenarios to evaluate the impact of the conventional tag's presence on the read rate of \SystemName. In all scenarios, we set \SystemName to read five \SystemSlave{s} in five distinct bands. We put two conventional tags within \SystemMaster's range for the `mild' and `medium' impact cases and four conventional tags for the `heavy' and `extreme' cases. The deployment settings are shown in Fig. \ref{exp:conventional_tag_impact_scenario}. To accommodate the potential occurrence of conventional tags, we manually set the Q value to ensure an adequate number of Aloha slots—specifically, 4 and 8 slots per query, respectively.

Fig. \ref{exp:conventional_tag_impact} shows the comparison of the read rate with or without the conventional tag's impact, as well as the misreads of these conventional tags. In all cases, \SystemName is little affected, as discussed in \S\ref{subsec:interference1}. Nevertheless, there exist certain misreads. This is because the five independent RFID sessions in the frequency domain may sometimes align and form a receivable downlink command for the conventional tags. When these tags start to backscatter their information, collisions happen with all five bands and decrease the \SystemName read rate. Even in the worst case, the read rate drop is less than 3\%. Further avoidance can be done by distinguishing these RN16s and ignoring them. In short, \SystemName is highly robust to conventional tags' impact.

\subsection{Cross-band Misreading}\label{subsec:eva_cross_band}
A reliable FDMA requires zero interference or misreading across bands. We now evaluate the cross-band reading of \SystemSlave{s}. We use an Impinj R2000 handheld RFID reader to read all five types \SystemSlave{s} across frequencies. Since cross-band reading is more likely to occur at close range and low bit rate, we set the reader at 33dBm EIRP power and read the tags within 1 meter using a 40kbps rate.

The upper figure in Fig. \ref{exp:cross_frequency} shows the read range of different \SystemSlave{s} across all frequencies, while the lower specifically illustrates the read rate of \SystemSlave F3 at different frequencies and distances. \revision{Misreading outside the operating band occurs only when both the frequency and distance are very close. However, when \SystemMaster is used for parallel reading, the frequencies of all bands are naturally separated by more than 5MHz. This separation minimizes misreading, even when tags are placed in close proximity. In conclusion,} as long as each carrier frequency in \SystemMaster remains within its designated operating band, tags in that band achieve maximum read rates, while cross-band misreading remains negligible. \revision{This also indicates that narrowing the band spacing is feasible within certain limits, allowing for more than five FDMA bands and greater spectrum utilization.}
\begin{figure*}[t]
    \centering
    \begin{minipage}[t]{0.3\textwidth}\centering
    \includegraphics[width=0.98\textwidth]{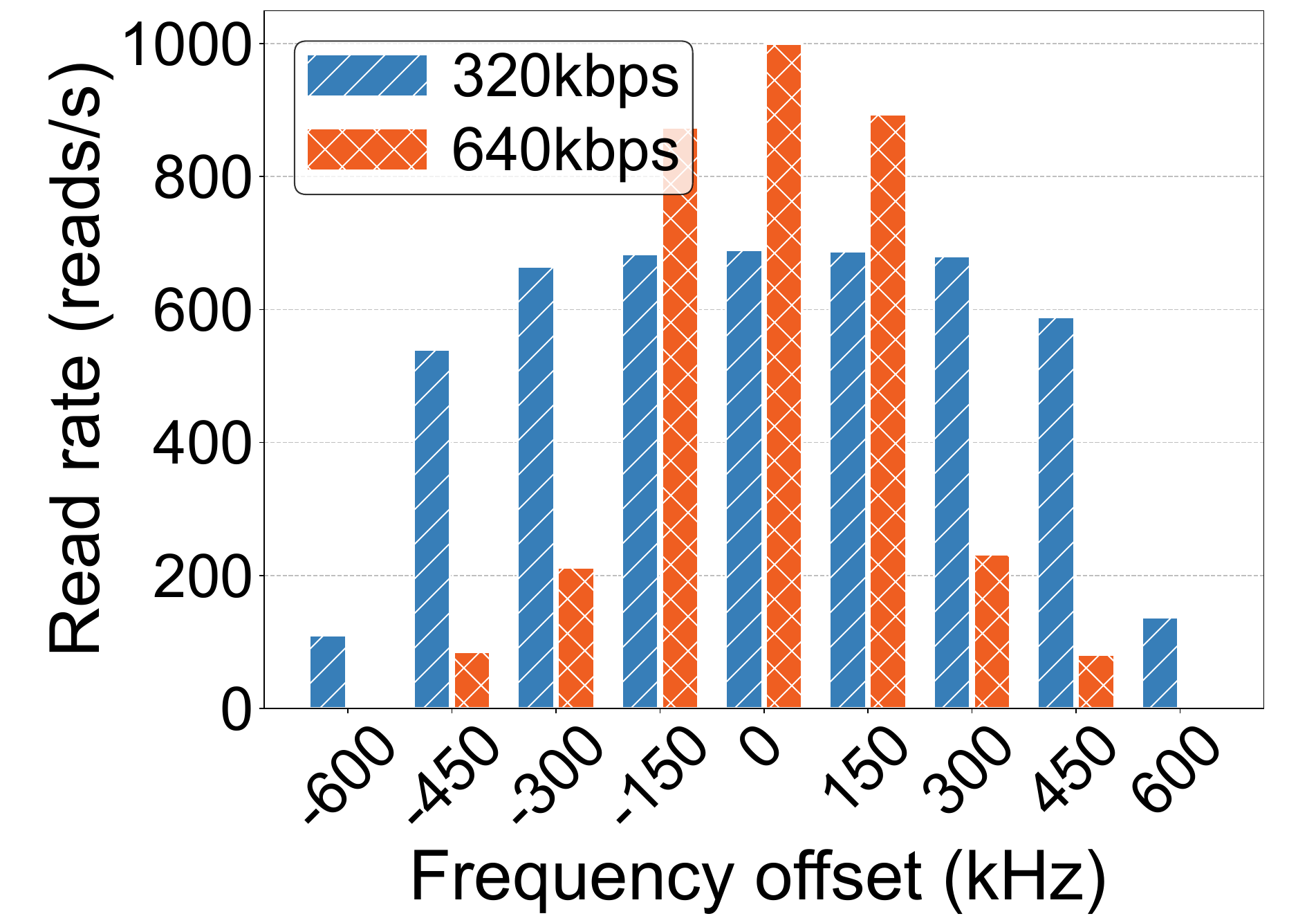}
    \vspace{-0.4cm}
    \caption{Impact of the carrier frequency offset.}\label{exp:datarate_frequency_offset}
    \end{minipage}
    \hspace{0.5cm}
    \begin{minipage}[t]{0.3\textwidth}\centering
    \includegraphics[width=0.98\textwidth]{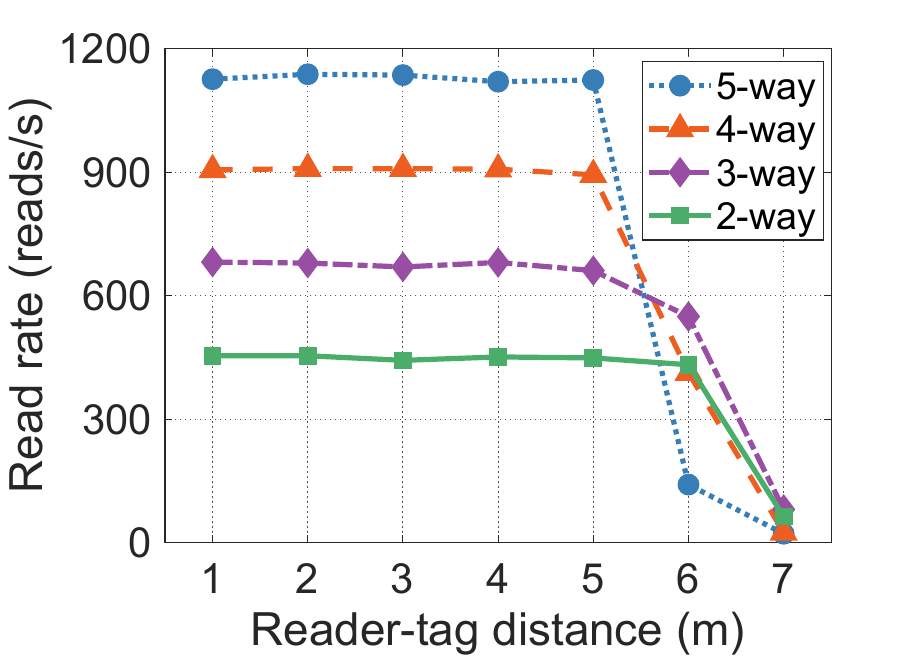}
    \vspace{-0.4cm}
    \caption{Trade off between parallelism and range.}\label{exp:nway_distance}
    \end{minipage}
    \hspace{0.5cm}
    \begin{minipage}[t]{0.3\textwidth}\centering
    \includegraphics[width=0.98\textwidth]{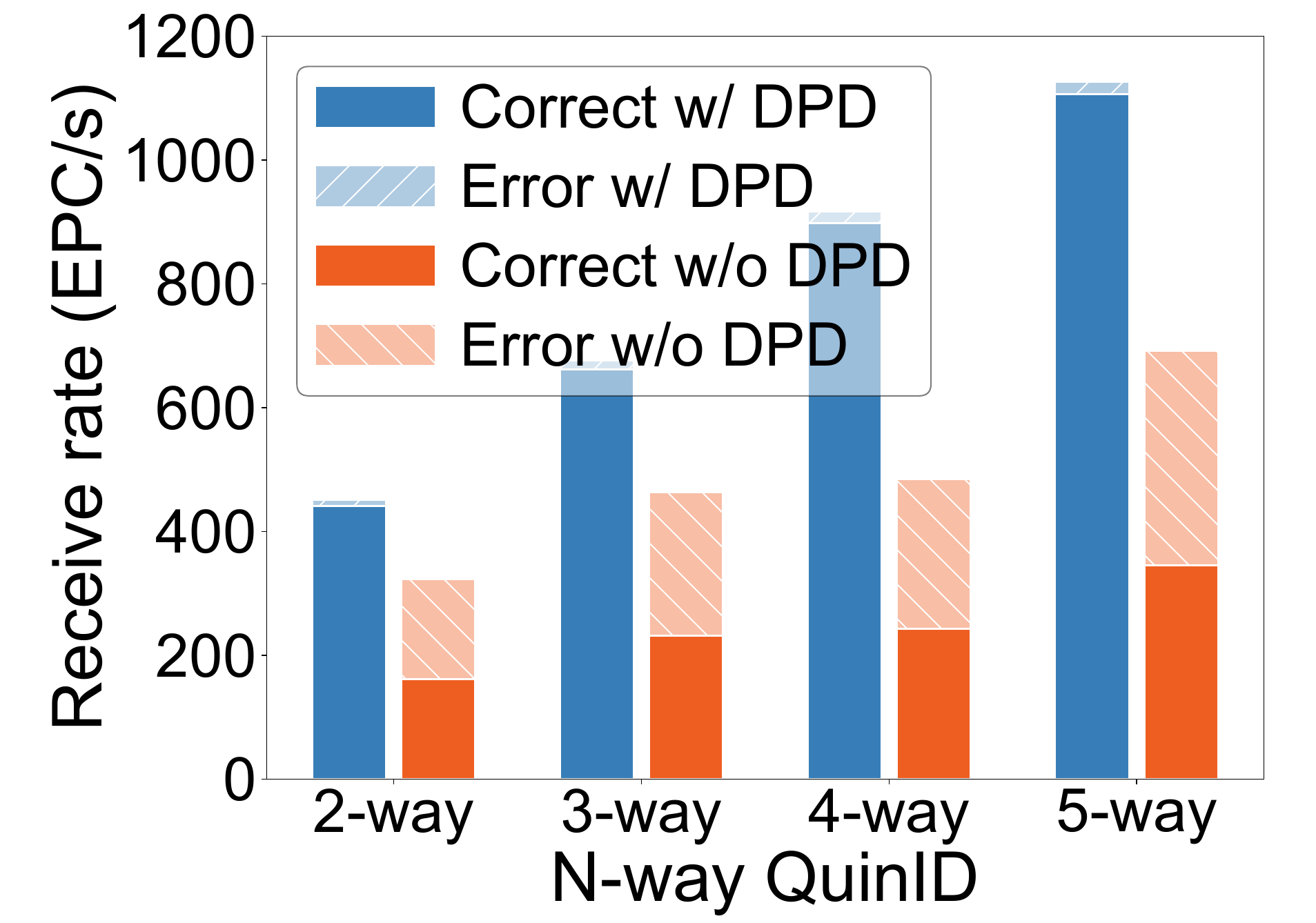}
    \vspace{-0.4cm}
    \caption{Effectiveness of the digital pre-distortion in \SystemMaster.}\label{exp:dpd_onoff_readrate}
    \end{minipage}
    \vspace{-0.5cm}
\end{figure*}
\subsection{Ablation Study}\label{subsec:ablation_study}
We conduct ablation studies to further assess \SystemName's performance. \S\ref{subsubsec:quintag_on_RFID_reader} evaluates \SystemSlave's performance on conventional readers and its compensation on the range reduction. \S\ref{subsubsec:eva_cfo} examines reader's carrier frequency offset effects. \S\ref{subsubsec:eva_parallelism} explores the trade off between range and read rate. \S\ref{subsubsec:dpd_evalute} evaluates the effectiveness of the digital pre-distorter.

\subsubsection{\SystemSlave with RFID reader} \label{subsubsec:quintag_on_RFID_reader}
\revision{This experiment compares \SystemSlave's performance to standard RFID tags when using commercial RFID readers. We use an Impinj R2000-based handheld reader, operating at 30dBm EIRP with a consistent 99\% modulation depth. \SystemSlave's PCB antenna, implemented with copper, has the same gain as the printed aluminum antenna in commercial tags, both featuring 2dBi directivity and 92\%-95\% efficiency \cite{AntennaEfficiency}.} Fig. \ref{exp:conventional_reader} compares the read rate at different distances. We can see that \SystemSlave performs similarly to standard RFID within the 6-meter range. The read rate decreases significantly at 7 meters.

By filtering out noise and improving downlink SNR, the SAW filter can partially offset its insertion loss, compensating for range reduction. We reduce the \revision{reader's} power by 3dB \revision{while maintaining a 99\% modulation depth} and read a commercial RFID tag, emulating \SystemSlave with SAW's insertion loss. \revision{This power reduction proportionally decreases the PIE-modulated signal, which represents the power difference between high and low states.} The resulting improvement confirms the analysis in \S\ref{subsec:range_reduction_compensation}, showing an acceptable communication range of \SystemSlave.

\subsubsection{Impact of carrier frequency offset}\label{subsubsec:eva_cfo}
Supporting high bit rates requires \SystemSlave to provide sufficient bandwidth. Fig. \ref{exp:datarate_spectrum} shows the operating bandwidth of a specific \SystemSlave and the FM0 encoded backscatter signal spectrum. We can see that the operating band covers the whole spectrum for both 320kbps and 640kbps bit rates. However, in practical reader implementations, the carrier frequency may not align perfectly with the center frequency of \SystemSlave. We evaluate the impact of possible carrier frequency offset. In this experiment, we vary the carrier frequency in a 150kHz step and measure the read rate under 320kbps and 640kbps bit rate. The results are shown in Fig. \ref{exp:datarate_frequency_offset}. We can see that when the offset is higher than a certain threshold, the performance decreases significantly. This is because the backscatter spectrum starts to fall outside the operating bandwidth with a high offset and gets filtered out by the SAW. The results show that \SystemSlave can tolerate about 150kHz offset at 640kbps bit rate and 450kHz offset at 320kbps. This level of carrier accuracy can be satisfied with modern RFID readers \cite{ST25RU3993}.
\subsubsection{Trade-off between parallelism and range}\label{subsubsec:eva_parallelism}
Our implementation includes five independent RFID sessions across different bands with equal transmission power distribution. Given a fixed total transmission power, \SystemName can trade off between read range and maximum read rate by adjusting the number of parallel sessions. We measure read rates at various distances in different session numbers, as shown in Fig. \ref{exp:nway_distance}. We find that the stable read distance can be extended to 6 meters when two parallel sessions are running, as the 2-way \SystemName increases the power by 4dB compared to the 5-way \SystemName. We leave the trade-off choices to users as they can flexibly select the parallelism of \SystemName according to their application needs.
\subsubsection{Effectiveness of digital pre-distortion}\label{subsubsec:dpd_evalute}
\SystemMaster includes a digital pre-distortion stage before amplifying the signal to cancel the inter-band interference. In this experiment, we show the effectiveness of this DPD stage. We switch the DPD on and off in different degree of parallelism configurations and measure the receiving rate of CRC-correct and CRC-error EPC packets. As shown in Fig. \ref{exp:dpd_onoff_readrate}, regardless of the degree of parallelism configured, inter-band interference exists as long as multi-band carriers are present in the frequency domain. This interference causes failure to detect the preamble of backscatter packets, resulting in fewer EPC packets received. It also corrupts the backscatter signal, causing bit decoding failure and CRC error in the EPC packets. With DPD enabled in \SystemName, the interference is successfully eliminated, and the read rate increases linearly with the increasing parallelism with few CRC errors.

% !TeX root = ../main.tex

%-------------------------------------------------------------------------------
\section{Related Work}\label{sec:related_work}
%-------------------------------------------------------------------------------
\subsection{Parallel Backscatter}
Backscatter is a promising solution for ubiquitous sensing in IoT applications \cite{RFinder, SensingFinger, RFID_Attendance_System}. Many efforts aim to boost its communication throughput \cite{Pushing_Throughput_OFDM_Backscatter, Mighty,RFTransformer_ToN}, particularly through parallelizing multiple tag transmissions. We categorize existing works into two groups: those collaborating with conventional OOK tags similar to RFID and those designing new tags employing diverse abilities.

\textbf{Parallel RFID.} These works adopt ``parallel decoding'', which involves decoding collided signals from parallel transmissions at the reader using time or IQ domain features \cite{Physical_Layer_Collision_Recovery, Laissez-Faire, Laissez-Faire-hotwireless, Come_and_Be_Served, DroneRelay, wangjue_Efficient2012}. At high sampling rates, collided OOK states form multiple clusters, with most works relying on the distinguishable ones for decoding. However, diverse tag channels and the exponential growth of clusters incur a significant superclustering phenomenon. To address this, recent works introduce additional information into the decoding, include temporal burstiness \cite{Hubble, Fireworks}, multi-frequency channel information \cite{Canon}, and multiple antennas \cite{FreeScatter}.

While these works allow parallel decoding of backscatter signals, the mandatory handshaking (exchanging an RN16) in the EPC protocol and their unstable performance significantly limit their practical usage. Their overall improvement in the read rate is minimal, amounting to less than 20\%. In comparison, \SystemName achieves fully parallel RFID reading and demonstrates a stable 5$\times$ increase in the read rate.

\textbf{New parallel backscatter designs.} In addition to decoding concurrent signals, in recent years, researchers aim to design new backscatter systems and enhance the capabilities of tags to achieve parallelism \cite{Hawkeye, Aloba, uTag}. Early approaches directly employ coding mechanisms on tags for collision recovery \cite{Parallel_Identification, Turbocharging}. These CDMA-based methods introduce prohibitive overhead in large-scale deployments. Subsequent efforts \cite{DigiScatter, NetScatter} propose using frequency shifting on multiple tags to create OFDMA signals, enabling extensive parallelism of up to 100 tags. Nevertheless, these methods often result in high energy overhead on the tags. Recent works introduce LoRa backscatter systems, capitalizing on chirp signals' features for parallel transmissions \cite{P2LoRa, Aloba_ToN}. Despite this, LoRa backscatter systems exhibit low data rates and are more suitable for long-range transmissions.

These new systems indeed offer potential for large-scale ID collection. However, they often demand more energy than an RFID tag can afford, which is typically only 1$\mu$W. Further, considering billions of RFID tags already in circulation and the widespread deployment of readers, their lack of compatibility inevitably causes resistance from the industry. In contrast, \SystemName supports fully parallel RFID reading. It is specifically designed to work in RFID applications and maintains compatibility with existing RFID infrastructure.

\subsection{SAW Technology in Battery-free IoT}
Due to their battery-free nature and ease of manufacturing, surface acoustic wave devices are extensively valuable for low-power IoT \cite{Saiyan_Ton}. They transduce physical signals into electrical ones, thus enabling sensing of pressure, temperature, and mass \cite{mandal2022surface, islam2012surface}. In \cite{Saiyan}, the authors employ SAW devices as differential circuits to achieve low-power LoRa signal demodulation. Further, leveraging the unique physics of surface acoustic waves, SAW devices can function as RFID tags \cite{Turcu_2009_SAWRFID}. They backscatter hard-coded ID information while achieving significant cost-effectiveness.

Different from these works, \SystemName is the first to integrate SAW filters into RFID antennas for frequency-domain parallelism. Although SAW devices have been employed directly as RFID tags, they do not support parallel operation.

% !TeX root = ../main.tex

%-------------------------------------------------------------------------------
\section{Discussion}\label{sec:discussion}
%-------------------------------------------------------------------------------
\subsection{Cost Analysis and Application Scope}
\revision{The cost of a commercial RFID tag includes the IC chip, antenna, and assembly. Using a flip-chip process, the typical cost is 3.4\textcent, with 1.3\textcent\ for the IC, 1\textcent\ for the antenna, and 1.1\textcent\ for assembly \cite{swamy2003manufacturing}. \SystemSlave adds a SAW filter, two inductors, and additional assembly. SAW filters, made with a simpler manufacturing process and cheaper piezoelectric materials \cite{mei2020fabrication}, are comparable in cost to or cheaper than RFID ICs \cite{SAW_cost_analysis}. Inductors, benefiting from a straightforward manufacturing process, cost about 1\textcent\ each. With an estimated fourfold increase in assembly cost, \SystemSlave is expected to cost under 10\textcent\ in mass production, about 3$\times$ that of a commercial tag.}

\revision{To further optimize costs, the RFID IC die can be integrated with the SAW filter die into a single module, with inductors implemented on the same IC. This significantly reduces assembly costs, with the primary increase concentrated in the module itself. A rough estimate places the module cost at around 3-4\textcent, making it 1.5-1.8$\times$ the cost of standard RFID tags. As for \SystemMaster, its cost is comparable to that of a standard RFID reader, with replacement costs incurred only at specific stages requiring parallel reading.}

\revision{Given these cost considerations, \SystemName is particularly well-suited for industrial applications such as large logistic centers, high-speed production lines, and valuable goods tracking. Although \SystemSlave has a relatively higher cost, the efficiency gains from improved performance often outweigh the additional expense. Moreover, the typically higher value of tracked items allows for gerater tolerance of the tag cost.}

\subsection{More Potential Applications}
Beyond parallel RFID, \SystemName can also enhance and expand other applications. In logistics, assigning tags of the same frequency band to each batch helps isolate batches, naturally preventing misreads during inventory replenishment. Additionally, tagging goods of the same category with the same band enables easy grouping of scattered items \cite{RFID_Grouping}.

\subsection{Reconfigurable \SystemSlave}
\revision{Currently, \SystemSlave{s} operate on preassigned frequency bands determined by the SAW filter, which may limit support for wideband techniques such as RFID localization. To enable reconfigurability, multiple SAW filters can be integrated with an RF switch for dynamic band selection. This would also enable channel hopping mechanisms on the tag, further improving reading efficiency. We believe such functionality could be incorporated into future RFID chips while maintaining compatibility with existing RFID systems.}

% !TeX root = ../main.tex

%-------------------------------------------------------------------------------
\section{Conclusion}\label{sec:conclusion}
%-------------------------------------------------------------------------------
We present \SystemName, an FMDA-based fully parallel RFID system. With a passive ultra-selective filtering antenna tailored for commercial RFID chips, tags in \SystemName achieve frequency-selective operation. Multiple independent RFID sessions are operating simultaneously across distinct frequency bands. \SystemName significantly increases the read rate of commercial RFID without significantly sacrificing read range and tag cost. From a boarder perspective, \SystemName introduces a passive RF computing technique \cite{Leggiero,Saiyan,RFTransformer} that operates directly on the frequency of RF signals. Our evaluation shows that the five band \SystemName implementation achieves a fivefold increase in the read rate, with a maximum of 5000 reads per second.

%----------------------------------- Acknowledgments -------------------------------
\begin{acks}
We sincerely thank our shepherd and reviewers for their valuable feedback. This work is supported in part by the National Natural Science Foundation of China under grant No. 62425207, No. 62472379, and No. 62272293.
\end{acks}

%----------------------------------- References ----------------------------------
\bibliographystyle{ACM-Reference-Format}
\bibliography{refs}

%----------------------------------- Appendix -------------------------------------
% \appendix

\end{document}